\documentclass[preprint]{aa}
\pdfoutput=1

\usepackage{amsmath}
\usepackage{amsfonts}
\usepackage{dsfont}
\usepackage{amsxtra}
\usepackage{hyperref}
\usepackage{amssymb}
\usepackage{upgreek}

\usepackage{multirow}
\usepackage{url}
\usepackage{float}

\usepackage{graphicx,epsfig}
\usepackage{dcolumn}

\usepackage{xspace} 

\usepackage{color}
\usepackage{xcolor}

\usepackage{lineno}
\newcommand{\be}{\begin{equation}}
\newcommand{\ee}{\end{equation}}
\newcommand{\bea}{\begin{eqnarray}}
\newcommand{\eea}{\end{eqnarray}}
\newcommand{\beaa}{\begin{eqnarray*}}
\newcommand{\eeaa}{\end{eqnarray*}}
\newcommand{\ba}{\begin{array}}
\newcommand{\ea}{\end{array}}
\newcommand{\bi}{\begin{itemize}}
\newcommand{\ei}{\end{itemize}}
\newcommand{\ben}{\begin{enumerate}}
\newcommand{\een}{\end{enumerate}}

\newcommand{\lb}{\label}

\newcommand{\adeg}{^\circ\!\!}

\newcommand{\Fermi}{\textit{Fermi}\xspace}
\newcommand{\pgw}{PGWave\xspace}

\newcommand{\onepic}{0.4}
\newcommand{\onepicabs}{0.35}
\newcommand{\mapscale}{0.17}

\definecolor{darkgreen}{rgb}{0.0, 0.7, 0.0}

\begin{document}

   \title{The first catalog of \Fermi-LAT sources below 100 MeV}

   %\subtitle{}

   \author{G.Principe \thanks{\email{giacomo.principe@fau.de}}
          \inst{1}
          \and
          D. Malyshev %\thanks{\email{dmitry.mayshev@fau.de}}
          \inst{1}
		  \and
          J. Ballet %\thanks{\email{jean.ballet@crea.fr}}
          \inst{2
          }
           \and
          S. Funk %\thanks{\email{s.funk@fau.de}}
          \inst{1
          }          
          }

   \institute{
             {Erlangen Centre for Astroparticle Physics, Erwin-Rommel-Str. 1, Erlangen, Germany}
              \and
               {Laboratoire AIM, CEA-IRFU/CNRS/Universit\'{e} Paris Diderot, Departement d’Astrophysique, CEA Saclay, Gif sur Yvette, France}
             }

   \date{Received March 27, 2018; accepted June 12, 2018}

% \abstract{}{}{}{}{} 
% 5 {} token are mandatory
 
\abstract{We present the first \Fermi Large Area Telescope (LAT) low energy catalog (1FLE) 
of sources detected in the energy range 30 -- 100 MeV.
The imaging Compton telescope (COMPTEL) onboard NASA's \textit{Compton Gamma-Ray Observatory} detected sources below 30 MeV, while
catalogs of point sources released by the \Fermi-LAT and EGRET collaborations use energies above 100 MeV.
Because the \Fermi LAT detects gamma rays with energies as low as 20 MeV,
we create a list of sources detected in the energy range between 30 and 100 MeV,
which closes a gap of point source analysis between the COMPTEL catalog and the \Fermi-LAT catalogs.
One of the main challenges in the analysis of point sources is the construction of the background diffuse emission model.
In our analysis, we use a background-independent method to search for point-like sources
based on a wavelet transform implemented in the PGWave code.
The 1FLE contains 198 sources detected above 3 $\sigma$ significance with eight years and nine months of the \Fermi-LAT data.
For 187 sources in the 1FLE catalog we have found an association in the \Fermi-LAT 3FGL catalog:
148 are extragalactic, 22 are Galactic, and 17 are unclassified in the 3FGL.
The ratio of the number of flat spectrum radio quasars (FSRQ) to BL Lacertae (BL Lacs) in 1FLE is three to one,
which can be compared with an approximately 1 to 1 ratio for the 3FGL or a one to six ratio for 3FHL.
The higher ratio of the FSRQs in the 1FLE is expected due to generally softer spectra of FSRQs
relative to BL Lacs.
Most BL Lacs in 1FLE are of low-synchrotron peaked blazar type (18 out of 31), which have softer spectra and higher redshifts
than BL Lacs on average.
Correspondingly, we find that the average redshift of the BL Lacs in 1FLE is higher than in 3FGL or 3FHL.
There are 11 sources that do not have associations in the 3FGL.
Most of the unassociated sources either come from regions of bright diffuse emission or 
have several known 3FGL sources in the vicinity, which can lead to source confusion.
The remaining unassociated sources have significance less than 4 $\sigma$.
}


\maketitle

%\tableofcontents

\setlength{\columnsep}{5mm}

\section{Introduction}
\lb{sec:intro}

% to introduce LAT and the LAT catalogs
The Large Area Telescope \citep[LAT,][]{2009ApJ...697.1071A} on board the \textit{Fermi Gamma-ray Space Telescope} has revolutionized our knowledge of the high-energy sky.
The LAT detects gamma-rays in the energy range from 20 MeV to more than 300 GeV, measuring their arrival times, energies, and directions.

% 3FGL catalog >100MeV
Although the LAT observations start at 20 MeV, all previous catalogs released by the \Fermi-LAT Collaboration are produced using optimized analysis focused on energies larger than 100 MeV.  
In particular, the Third \Fermi-LAT catalog \citep[3FGL,][]{2015ApJS..218...23A} characterizes 3033 sources in the energy range between 100 MeV and 300 GeV from the first four years of LAT data. 
Since the sensitivity of the instrument peaks at about 1 GeV, the 3FGL favors sources that are brightest around these energies.
At energies below 100 MeV the analysis of point sources is complicated due to large uncertainties in the arrival directions of the gamma rays, which leads to confusion among point sources, difficulties in separating point sources from diffuse emission, and high contamination from the Earth limb. 
In this paper, we have used a background-independent analysis of point sources based on a wavelet transform of the gamma-ray data, which filters out the large-scale diffuse emission and the Earth limb contamination. 

The EGRET telescope \citep{1992NASCP3137..116H}, which is a preceding gamma-ray experiment, measured gamma-rays from 20 MeV to 30 GeV. However, the catalogs released by the EGRET collaboration only used data above 100 MeV \citep[e.g.,][]{1999ApJS..123...79H}. At lower energies, COMPTEL analyzed the gamma-ray sky between 0.75 and 30 MeV \citep{2000A&AS..143..145S}.
Therefore, the energy range from 30 MeV to 100 MeV was not covered by any of the previous gamma-ray point source (PS) catalog analyses.

% The 1FLE catalog
In this paper, we present the first catalog of sources detected from 30 MeV to 100 MeV by the \Fermi LAT. The first \textit{Fermi} Low Energy catalog (1FLE) is constructed using 8.7 years of LAT data taking full advantage of the improvements provided by the Pass 8 data selection and event-level analysis \citep{2013arXiv1303.3514A}, in particular the large increase of acceptance at low energy ($>$70\% below 100 MeV) and the point-spread-function (PSF) event type classification\footnote{A measure of the quality of the direction reconstruction is used to assign events to four quartiles}. 
Special attention is given to the different PSF event type selections, in particular to the data cuts used to maximize the detection rate in the 1FLE.

% Method used - PGWave a bkg indipendent method
%Since one of the largest uncertainties in the PS studies below 100 MeV is the uncertainty in the diffuse background, 
In this analysis we use the wavelet transform method implemented in the PGWave tool \citep{1997ApJ...483..350D}.
PGWave is already used in the \Fermi-LAT catalog pipeline as one of the methods to find PS candidates (so-called seeds).
In the standard catalog pipeline (e.g., for the 3FGL), the PS candidates are further evaluated with a likelihood analysis to refine the positions of the sources and to determine the fluxes.
In contrast, in this paper, we have used the wavelet transform both to detect the sources and to estimate their fluxes
\citep[see, e.g., ][for a discussion of the flux determination with the wavelet transformation]{2016arXiv161001351P}.

%Description of the Chapter
The paper is structured as follows: in Section 2 we describe the \Fermi-LAT data used in the analysis. 
A description of the analysis, in particular of the reconstruction of PS position and flux with the PGWave tool, is provided in Section 3. 
In Section 4 we present the 1FLE catalog and compare it with 3FGL and COMPTEL catalogs. 
Section 5 contains the conclusions.

%\end{multicols}

%\begin{multicols}{2}

\section{Data selection}
\lb{sec:data}

We used eight years and nine months of the \Fermi-LAT Pass 8 Source class events, with the P8R2\_SOURCE\_v6 instrument response functions (IRFs), 
between August 4, 2008  and May 3, 2017 ({\Fermi} Mission Elapsed Time 239557418\,s--515548139\,s).
For the wavelet analysis we used logarithmic energy bins with two bins per decade: 31.6  -- 100 MeV and 100 -- 316 MeV.
%In the analysis we use the energy range from 31.6($10^{1.5}$) to 316($10^{2.5}$) MeV for even spacing in logarithmic scale.
The second energy bin, 100 -- 316 MeV, is used for a crosscheck with the 3FGL.
For simplicity, we use the notation 30 -- 100 MeV and 100 -- 300 MeV when we refer to the energy bins in the following.
In order to reduce contamination from cosmic-ray interactions in the Earth atmosphere, 
we select events with an angle $\theta < 90\adeg$ with respect to the local zenith.

\begin{figure*}[h]
\centering
\includegraphics[scale=0.46]{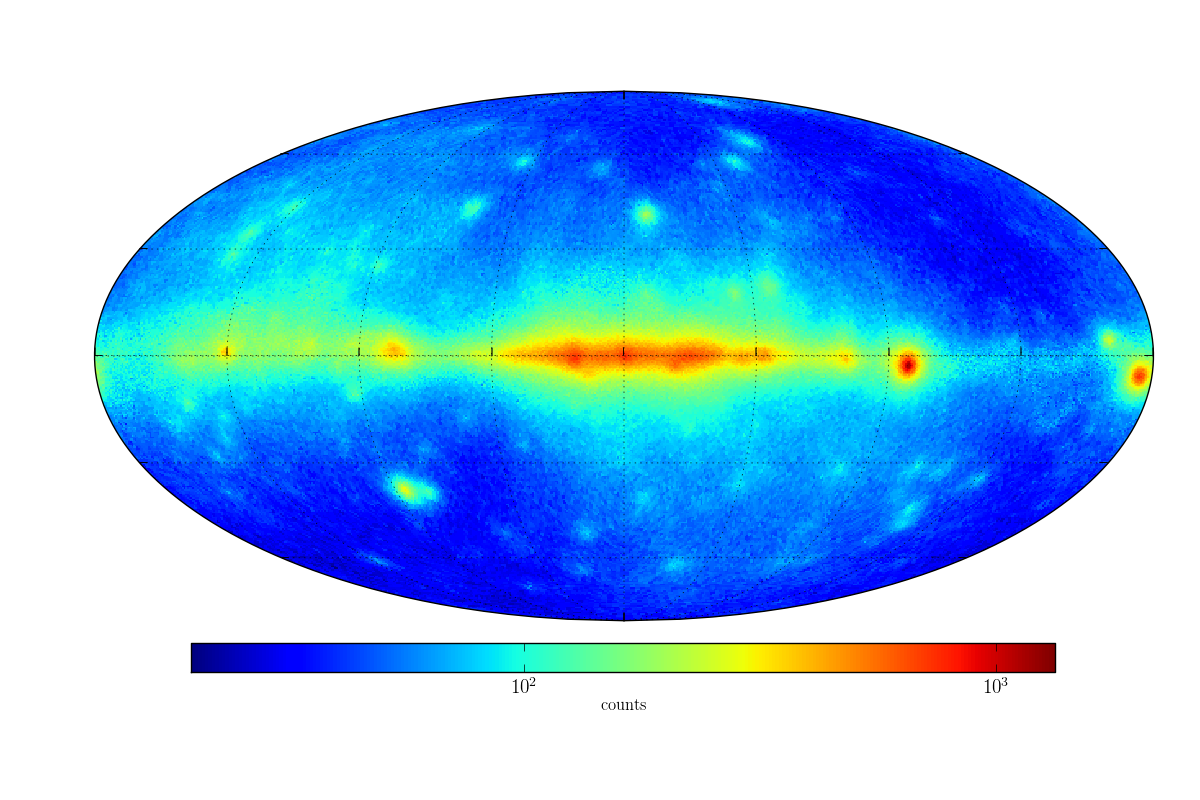}
\caption{\small \label{counts_map}
\Fermi-LAT count map, for the first eight years and nine months of observation, in the 30 -- 100 MeV band represented in Galactic coordinates and Mollweide projection. The color scale is logarithmic and the units are counts per pixel (pixel size equal to 0\fdg458).
}
\end{figure*}

Gamma rays in Pass 8 data can be separated into 4 PSF event types: 0, 1, 2, 3, where PSF0 has the largest point spread function and PSF3 has the best.
We tested the performance of the PS detection algorithm for different PSF event types (see Appendix \ref{sec:PSF_selection}).
We considered the following combinations of PSF event types:
\bi
\item
all PSF event types combined together;
\item
PSF1, PSF2 and PSF3 event types; 
\item
PSF2 and PSF3 event types;
\item
only PSF3 event type.
\ei
We find that using only PSF3 event type gives the highest detection efficiency and the smallest false positive rate (see Appendix \ref{appendix_mc_simulations}).
Consequently, we used the PSF3 event type in our analysis.
Figure \ref{counts_map} shows the counts map for the first eight years and nine months of the \Fermi-LAT data between 30 and 100 MeV with PSF3 event type.

%\onecolumn
%\end{multicols}

%\clearpage
%\twocolumn

%\begin{multicols}{2}

\section{Point source analysis}
In this section we have used simulated gamma-ray maps to optimize the parameters in our PS detection algorithm and 
to determine the localization and flux reconstruction uncertainties.

\subsection{Source detection}
A common problem for the analysis of point sources in the \Fermi-LAT data is to determine an accurate Galactic diffuse emission model. 
If the diffuse model does not correctly reproduce the background of the sky, it can introduce large systematic errors in the PS detection and in the flux determination. 
In order to be less sensitive to the choice of the background model, we used a background-independent analysis method based on the wavelet transform
implemented in the PGWave tool (for more information see Appendix \ref{pgwave}).
We set the threshold for detection to 3$\sigma$ statistical significance. 
We used Monte Carlo (MC) simulations to optimize parameters, such as the wavelet transform scale and the minimum distance between two sources,
in order to maximize the detection efficiency and minimize the false positive rate
(see Appendix \ref{sec:PGWparam_selection}).

\subsubsection{Analysis procedure}
\label{analysis_procedure}
%ROI
All data, diffuse background, and MC maps are generated using HEALPix\footnote{\url{http://sourceforge.net/projects/healpix/}}
 \citep{2005ApJ...622..759G} 
pixellation with $n_{side}$ = 128 (pixel size$=\sqrt{\frac{4 \pi}{N_{pixel}}} \approx 0\adeg.458$, with $N_{pixel}=12 n_{side}^{2}$).
For the analysis, we projected the HEALPix maps to 12 regions of interest (ROIs) of size 180$\adeg$ in longitude and 90$\adeg$ in latitude
using Mollweide (MOL) projection \citep{2002A&A...395.1077C}.
The pixel size near the center of the ROIs is chosen to be 0\fdg458, so the HEALPix data are mapped as closely as possible to the MOL projection.
The centers of ROIs are at $b = 0\adeg,\; \pm 60\adeg$ and $\ell = 0\adeg,\; 90\adeg,\; 180\adeg,\; 270\adeg$.
The centers of the ROIs are chosen in order to cover the entire sky, to have an overlap between nearby regions and to have a maximum distance between a considered point in the ROI and the local equator, namely the local x-axis in the center of the ROI, less than $30\adeg$.
We studied the dependency of the detection efficiency varying the distance from the local equator and we do not see any relevant difference: the projection used does not affect the detection rate. 

For each ROI, we performed the wavelet transform with the PGWave tool and eliminate the seeds that are closer to the border than 10$\adeg$ to avoid edge effects. 
We merged the seeds in the overlapping regions between different ROIs, eliminating duplicates within 2$\adeg$ and giving a preference to the projection in which the source is closest to the equator.

In order to avoid contamination due to the diffuse emission, we repeated the previous steps of the analysis also for simulated maps of the Galactic and extra-galactic diffuse emission only. 
In our list we eliminated the seeds that match those in the purely diffuse maps.
Thus, a diffuse emission model (see Appendix	 \ref{appendix_mc_simulations}) enters in the analysis indirectly: we eliminated point-like features in the diffuse emission 
from the list of the PS seeds.
A comparison with the official diffuse emission model, provided by the Fermi-LAT collaboration \citep{2016ApJS..223...26A}, indicates that the two models are compatible and their difference results, on average, in one additional  false positive point source.

Finally, we estimated the flux of the sources using the maxima in the wavelet transform (WT) map (referred to as WT peak values in the following for conciseness).
Diffuse emission can also affect the determination of the flux by introducing fluctuations of the wavelet transform map on top of the Poisson noise from the sources.
We evaluated this effect using MC simulations by comparing the input and the reconstructed fluxes (see Section \ref{flux_det}).

We also used MC simulations to find the optimal wavelet transform radii, which give a high detection efficiency while keeping the false positive rate at a few per cent level at high latitudes (see Appendix \ref{sec:PGWparam_selection}).
For the search of PS in the data, we combined the results of wavelet transforms with two radii: 1\fdg4 and 1\fdg8.
For radii smaller than 1\fdg4, the 68\% containment radius at 100 MeV is much larger than the wavelet transform scale;
as a result, the efficiency of detection of faint sources becomes smaller.
For radii larger than 1\fdg8, we start to lose sensitivity due to source confusion.
In our analysis, the wavelet transform with the 1\fdg8 scale is important to detect faint sources at high latitudes,
while the transform with the 1\fdg4 radius helps to improve source confusion, especially close to the Galactic plane.
In the analysis below, we have combined the results of analysis with 1\fdg4 and 1\fdg8 wavelet scales. We gave the preference to the results of the 1\fdg8 wavelet scale analysis: if a source is found in both the analyses, we used the position and the WT peak of the 1\fdg8 wavelet scale analysis since it is more sensitive to faint sources.

\subsubsection{Detection efficiency}
\lb{sec:det_eff}
For the determination of the detection efficiency, we used MC maps that include diffuse emission and 3033 PS (obtained by extrapolating the 3FGL sources) with random positions in the sky 
(details on simulated maps can be found in Appendix \ref{appendix_mc_simulations}).
% Comparison
We applied the analysis to the simulated maps and we compare the resulting sources with the list of input sources in order to estimate the detection efficiency and the false positive rate, 
which are the ratio between the number of detected sources and the input sources and the ratio between the PGWave seeds that do not have an association and the total number of seeds.

\begin{figure}[h]
\centering
%\hspace{-1cm}
\includegraphics[scale=\onepic]{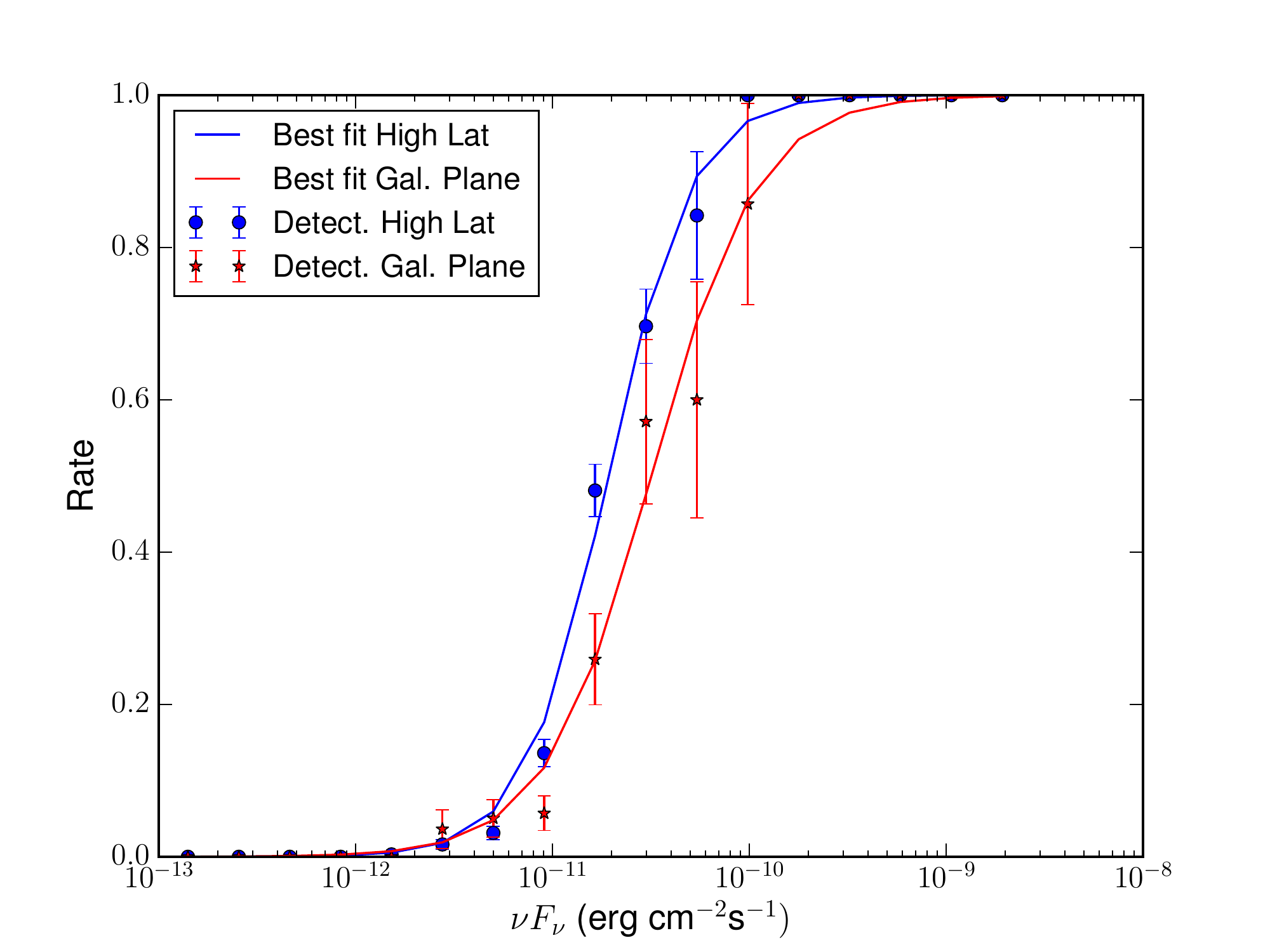}
\caption{\small \label{detection_flux}
Detection rate as a function of the input flux of the simulated PS with random position in the sky (setup containing extrapolated 3FGL sources with randomized position, Appendix \ref{appendix_mc_simulations}). 
Each point represents the ratio of detected sources using PGWave with respect to the number of input sources with a flux value inside the flux bin. The error bars corresponds to the statistical uncertainty derived with the binomial law.
%They are binned in flux of the input PS.
}
\end{figure}

%Results comparison with Input MC
Figure \ref{detection_flux} shows the ratio of detected sources to the input MC sources in each of the flux bins for high latitude sources, 
namely $\mid b \mid > 10 \adeg$, and Galactic sources.
%The plots show that PGWave is sensitive only to sources with a flux above $\sim 10^{-6}{\rm \; cm^{-2}s^{-1}}$.
We modeled the detection efficiency by a hyperbolic tangent function 
$({1 + \tanh\lambda(f - f_0)})/{2},$
where parameters $\lambda$ and $f_0$ are 
determined by fitting to the detection efficiency points.
Using this model at $\mid b \mid > 10 \adeg$, and the obtained values of $\lambda =1.079 $ and $f_0 = 1.876 \times 10^{-11}{\rm \; erg \; cm^{-2}s^{-1}}$ ($\chi^{2}=0.24$), 
we find that PGWave has a detection rate larger than 95\% for PS with $\nu F_{\nu}= {\rm Energy\;Flux} / \ln(E_{max}/E_{min})>6.5 \times 10^{-11}{\rm \; erg \; cm^{-2}s^{-1}}$ at 56 MeV.

\subsection{Localization}
The position of each source was first determined with PGWave, as the position of the WT maximum. Since PGWave returns the positions of the center of the pixel in which the WT has a maximum, we optimized the reconstruction of the position using a parabolic fit in a $5 \times 5$ pixel grid around the maximum (see Appendix \ref{Optimized_Localization}).
To determine the statistical and systematic uncertainties of the localization we made ten realizations of simulated maps.
Each simulated map contains 369 PS randomly positioned in the sky (see Appendix \ref{appendix_mc_simulations}). 
 
% Statistical Unc
In order to determine the statistical uncertainty, we estimated for each PS ($k$) the 1D dispersion of the reconstructed PS positions relative to the average position (in 2D) of the source in the MC realizations:

\begin{equation}
\sigma_{k} = \sqrt{\frac{\Sigma_{n} (X_{PGW_{i}}-X_{PGW_{mean}})^{2}}{2(n-1)}},
\end{equation}
where $n = 10$ is the number of MC realizations, $X_{PGW_{i}}$ and $X_{PGW_{mean}}$ are the coordinates of the reconstructed position in the $i$-realization and of the average reconstructed position over all the realizations. 
%In this way we have a value of the statistic uncertainty due by the localization with PGWave. 
Then we averaged the dispersion among all the detected sources in five flux bins:

\begin{equation}
\sigma_{stat} = \sqrt{\frac{\Sigma_{k}\sigma_{k}^{2}}{N}},
\end{equation}
where $N$ is the number of sources in a flux bin.

\begin{figure}[h]
\centering
\includegraphics[scale=\onepic]{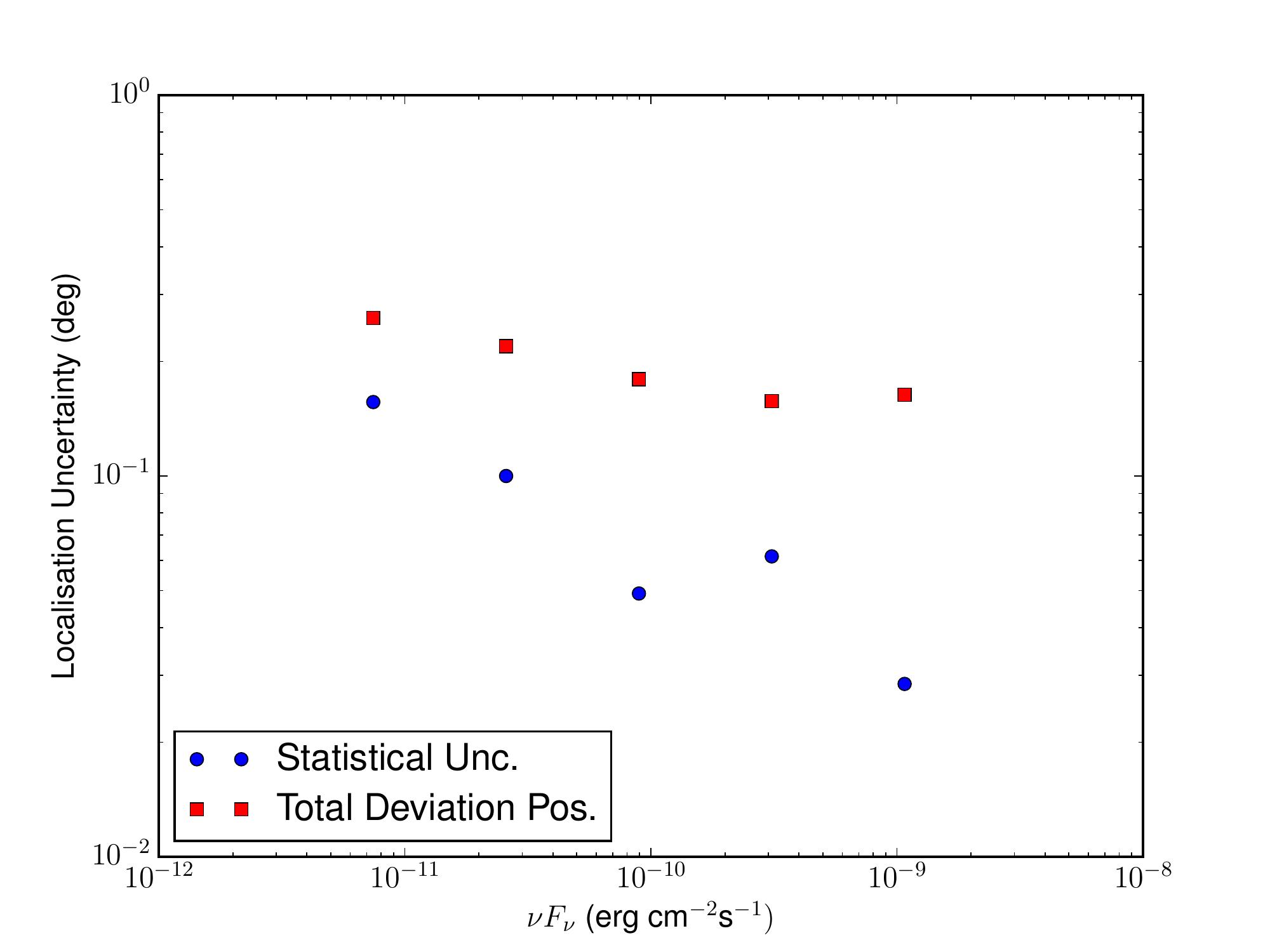}
\caption{\small \label{error_localization_30}
Statistical (blue circles) and systematic (red squares) uncertainty of the PGWave localization for the energy bin 30 -- 100 MeV. The uncertainties are reported as the 68\% containment radius.}
\end{figure}

%Systematic Unc.
For the systematic uncertainty, we measured the deviation between the averaged position of ten realizations and the input position of the PS:

\begin{equation}
\sigma_{syst} = \sqrt{\frac{\Sigma_{k} (X_{PGW_{mean}}-X_{in})^{2}}{2N}},
\end{equation}
where $X_{in}$ is the input position of the reconstructed source.
% Staistical and systematic Unc Localization
Figure \ref{error_localization_30} shows the statistical and systematic uncertainty in the localization. The total error in the localization is dominated by the systematic uncertainty. 
The value of the systematic uncertainty is smaller than 0\fdg3 in the energy range 30 -- 100 MeV and smaller than 0\fdg2 in the energy range 100 -- 300 MeV.

\subsection{Flux determination}
\label{flux_det}
\begin{figure}[h]
\centering
\includegraphics[scale=\onepic]{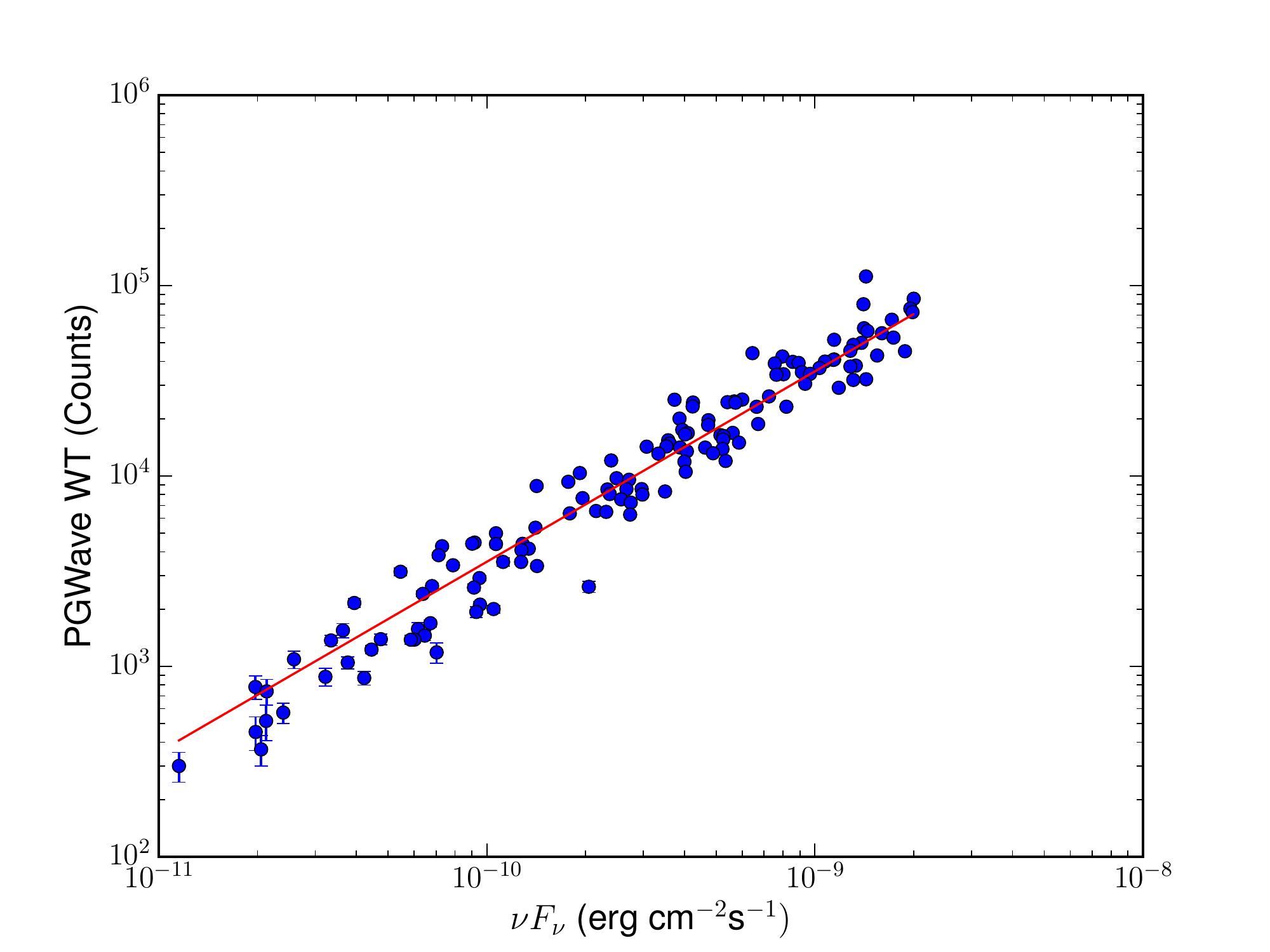}
\caption{\small \label{flux_reconstrucation_30}
PGWave seeds associated with the input PS in the energy bin 30 -- 100 MeV. X-axis: MC input flux, y-axis: the WT peak value from PGWave. The red line represents the best fit.
The relation is plotted for the wavelet transform with radius 1\fdg8.
}
\end{figure}

For the estimation of the flux of sources detected by PGWave, we used the WT peak value, 
because a linear correlation is expected between the flux and the WT peak value \citep{2016arXiv161001351P}.
In order to compare the results with the 3FGL catalog, 
we analyzed the sources also for the energy bin 100 -- 300 MeV using the same procedure as in the 30 -- 100 MeV bin.
Figure \ref{flux_reconstrucation_30} shows the correlation between the input MC flux and the WT peak values. 
The red line in the plot represents the best fit with a power-law function $f = k x^{\alpha}$, 
where $\alpha = 1$, which is expected by the definition of the WT (Appendix \ref{pgwave}).
%a value equal to 1 is expected for the power law index.
%For the power law index a value equal to 1 is expected, by the definition of the WT (see Appendix \ref{pgwave}). 
We compare a model with the fixed power law index $\alpha = 1$ to a model with $\alpha$ fitted to the data
(best-fit value $\alpha = 1.02$).
The difference between the two models in the reconstructed flux is less than 5\%, which is much
smaller than our systematic uncertainties (see below).
As a result, we used the simpler model with $\alpha = 1$.
%Since the differences  in the results between the two models are less than 5\%, much smaller than our systematic (see below), we have decided to use the model with a fixed power law index.
%Since the obtained values of the power law index are compatible with 1, we fix it in the best fit evaluation.
The power law factor $f$ is equal to $3.54 \times 10^{13}$, for the energy range between 30 and 100 MeV, and equal to $7.18 \times 10^{13}$, for the energy range between 100 and 300 MeV.
We made use of the results of the best fit for estimating the flux from the WT peak values.

\begin{figure}[h]
\centering
\includegraphics[scale=\onepic]{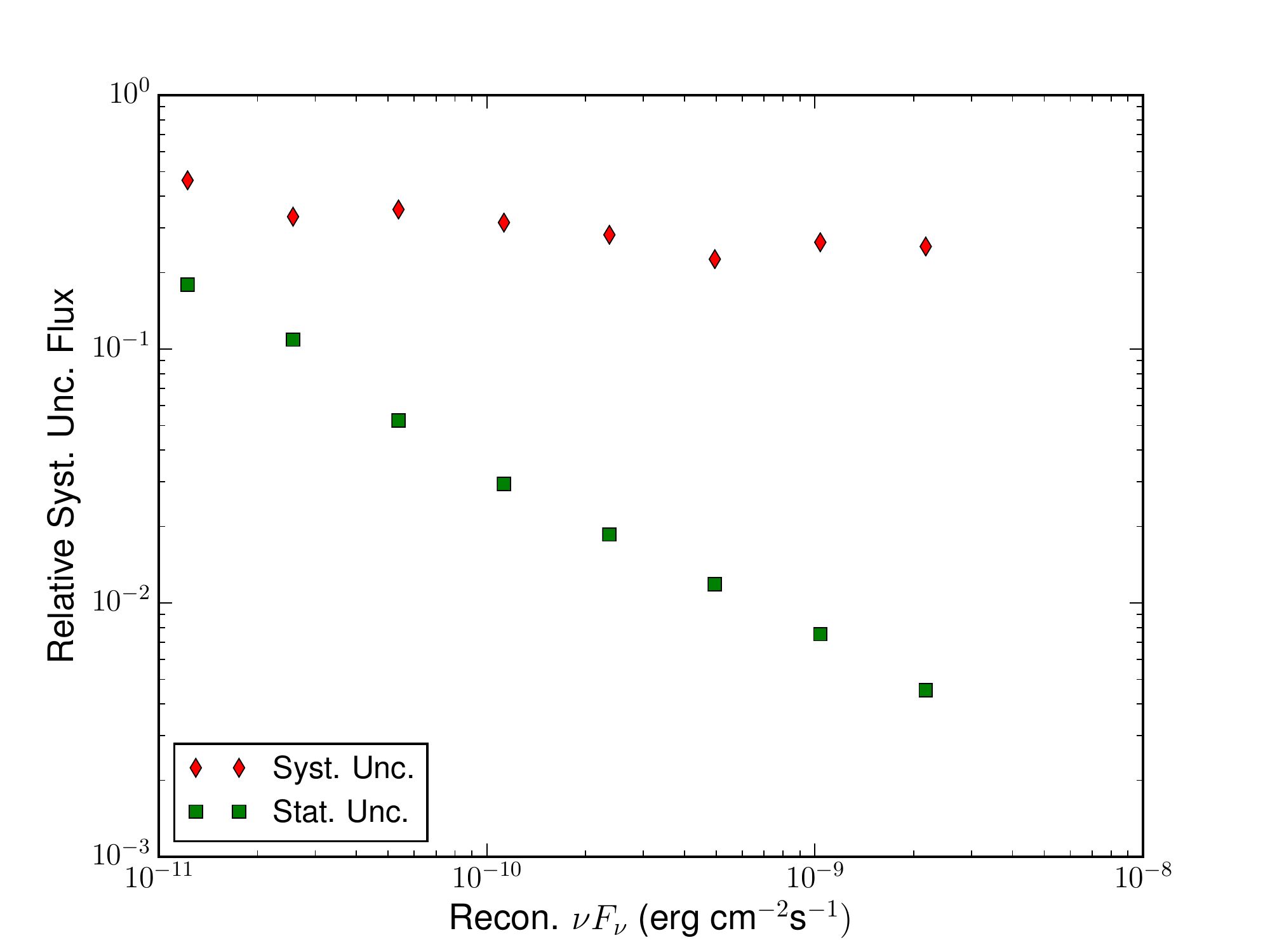}
\caption{\small \label{err_flux_reconstruction_30}
Statistical (green points) and systematic (red points) uncertainty of flux reconstruction using PGWave
in the energy range 30 -- 100 MeV.}
\end{figure}

%Statistical and Systematic Unc Flux reconstruction
To derive the systematic uncertainty of flux reconstruction, we divided the sources in bins of WT peak values, 
then we calculate inside each bin the standard deviation of the difference between the input MC flux and the PGWave best fit. 	
The statistical uncertainty is calculated by PGWave.
Figure \ref{err_flux_reconstruction_30} shows the systematic and statistical uncertainties in the flux reconstruction. 
In our case, the uncertainty is dominated by the systematic one. 
In the energy range 30 -- 100 MeV, the error decreases as a function of flux from a relative value of $\sim 40$\%, for faint sources, to a value of $\sim 20$\%, for bright sources. Similarly the error in the energy range 100 -- 300 MeV decreases as a function of the flux from a relative value of 35\%, for faint sources, to a value of 10\%, for bright sources.

\subsection{Association of point sources}
\label{association_algorithm}
The algorithm of PS association is based on positional coincidence with a tolerance radius of 1\fdg5, 
which is smaller than the angular resolution of PSF3 event type at 100 MeV ($3\adeg \,$), and on a flux ordering. 
The tolerance radius was chosen as the distance at which 98\% of the reconstructed sources find the correct associated input sources (see Appendix \ref{Optimized_Localization}).
The flux ordering algorithm associates seeds with a large WT peak value to bright MC input sources 
(or bright PS in a catalog when compared, e.g., to the 3FGL \Fermi-LAT catalog or the catalog of COMPTEL sources below), 
this allows a better association in cases when more than one source is present inside the tolerance radius.

%\end{multicols}

\section{The 1FLE Catalog}
% Introduction
In this section, we present the results of our point source analysis using 8.7 years of \textit{Fermi}-LAT data between 30 -- 100 MeV.
We applied the analysis described in Section \ref{analysis_procedure} to the data. 
We also compare the 1FLE catalog with the \textit{Fermi}-LAT 3FGL catalog derived above 100 MeV \citep{2015ApJS..218...23A} and 
with the COMPTEL catalog of sources at energies between 1 and 30 MeV \citep{2000A&AS..143..145S}.

\subsection{General characteristic of 1FLE sources}
The 1FLE catalog includes 198 sources detected over the whole sky (for a source detection we require a statistical confidence of more than 3$\sigma$)
that are not associated with significant seeds in the \pgw transform of the purely diffuse model map.
With the optimized parameters for the analysis (see Appendix \ref{sec:PGWparam_selection}) the expected number of spurious sources is about five.
The list of sources contained in the 1FLE is available only in electronic format (FITS format) as supplementary material. The columns are described in Table \ref{fits_file_table}.

\begin{figure}[h]
\centering
\includegraphics[scale=\onepic]{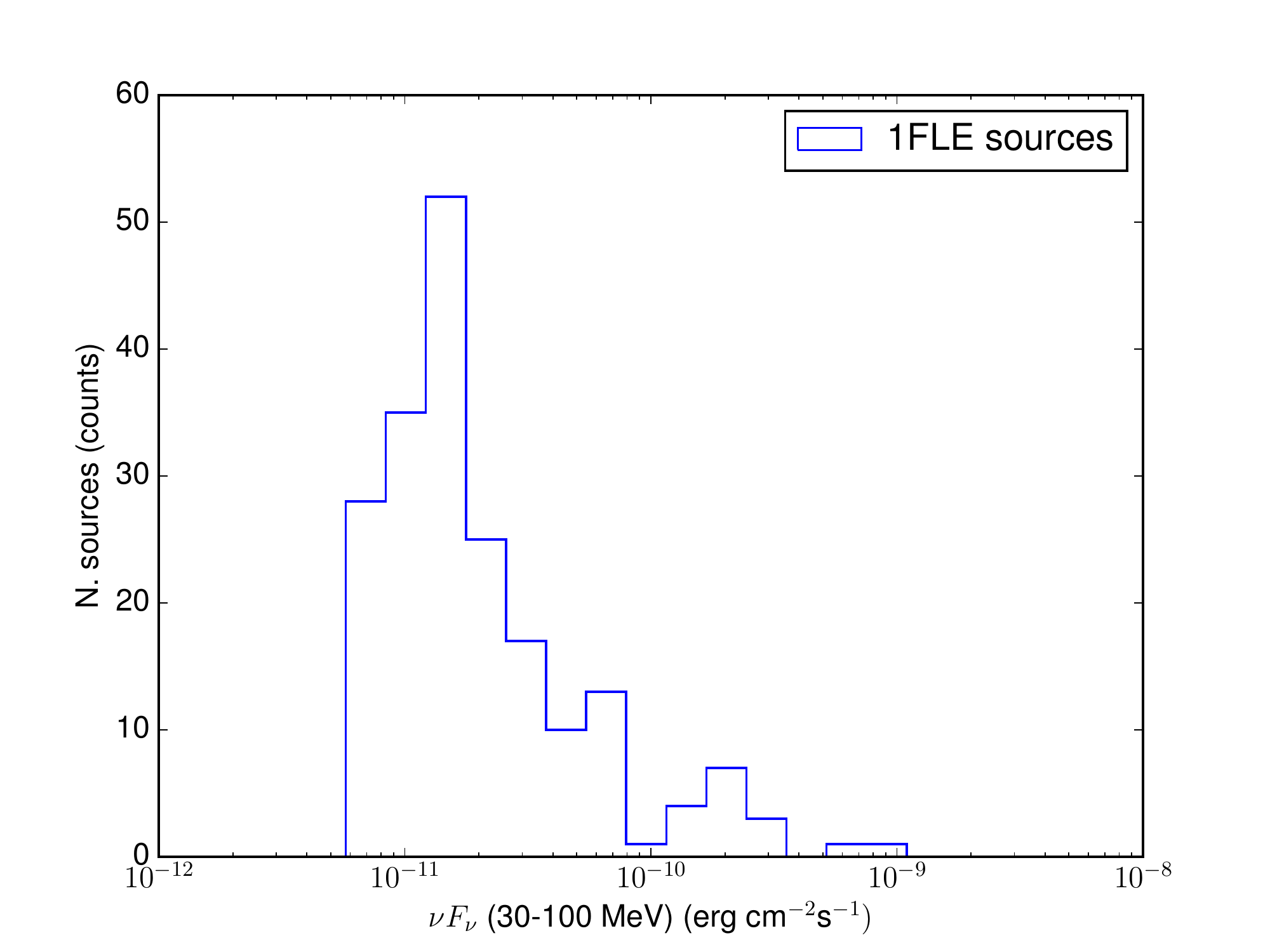}
\caption{\small \label{flux_1fle} 
Histogram of the flux between 30 and 100 MeV of the sources in 1FLE.}
\end{figure}

% Characteristics sources: Flux
Figure \ref{flux_1fle} shows the distribution of the flux between 30 and 100 MeV of the sources contained in the 1FLE.

The 1FLE sources have fluxes ($\nu F_{\nu}$) between 30 and 100 MeV in the range from $7 \times 10^{-12}$ erg cm$^{-2}$ s$^{-1}$ to $10^{-9}$ erg cm$^{-2}$ s$^{-1}$. 
The two brightest sources are the Vela pulsar and Crab pulsar and pulsar wind nebula.
Since we have not looked for the pulsed emission from the Crab pulsar, we cannot separate the flux from the pulsar itself and the pulsar wind nebula.
In the following, when we talk about the Crab pulsar, we assume the combined emission from the pulsar and from the pulsar wind nebula.

\begin{figure*}[h]
\includegraphics[scale=0.8]{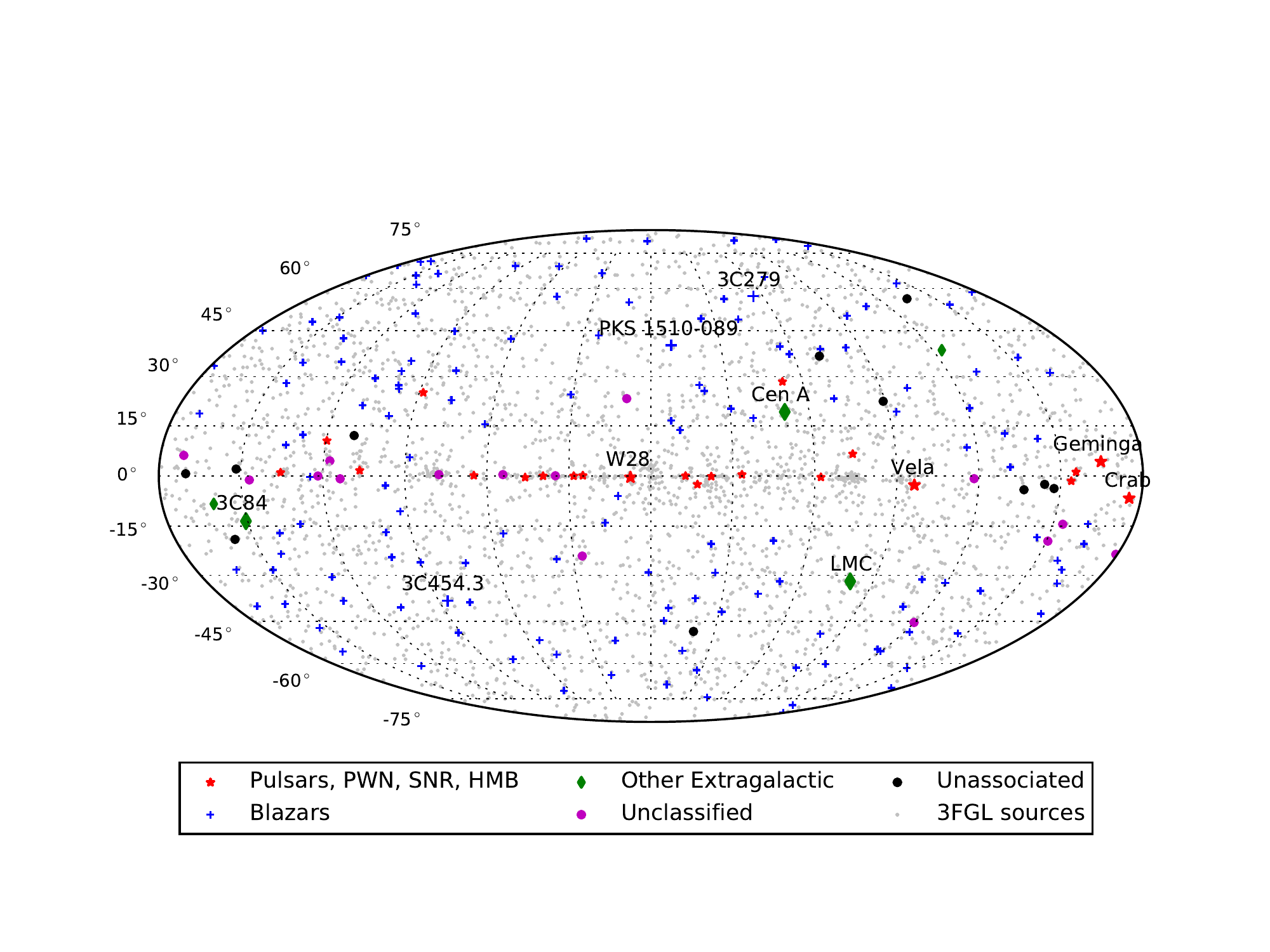}
\caption{\small \label{1fle_ps}
Sky map, in Galactic coordinates and Mollweide projection, showing the sources in the 1FLE catalog classified by their most likely association, using the 3FGL association. All the 3FGL sources are also plotted, with gray points, for a comparison.
}
\end{figure*}

% Characteristics sources: Location
Figure \ref{1fle_ps} shows the locations of the 1FLE sources in the sky. Most of the 1FLE sources (157) are at high latitude ($\mid b \mid >10\adeg$); only 41 are at low latitude.
%Respect a likelihood method, in which a source can be detected also on top of peaks of the diffuse emission, 
Our method is less sensitive to sources in the Galactic plane than likelihood methods, 
because we reject the seeds that match those from the purely diffuse emission map (see Section \ref{analysis_procedure}). 
This results in the exclusion of about 6\% of pixels within $|b| < 10\degr$ 
due to masking of $1\fdg5$ circles around 60 seeds found in the purely diffuse model.

%\clearpage
%\onecolumn

\begin{table*}[h]
\small
\centering
\begin{tabular}{cccc}
\hline
\hline
Column &Format & Unit & Description \\
\hline
Source\_Name & 18A & --- & Official source name 1FLE JHHMM+DDMM\\ 
RAJ2000 & E & deg & Right ascension \\
DEJ2000 & E & deg & Declination\\
GLON & E & deg & Galactic longitude\\
GLAT & E & deg & Galactic latitude \\
Conf\_95\_Radius & E & deg & Error radius at 95\% confidence \\
Signif\_Avg & E & --- & Source significance in $\sigma$ units over the 30 MeV to 100 MeV band \\
Energy\_Flux30\_100 & E & erg cm$^{-2}$ s$^{-1}$ & Energy flux ($\nu F_{\nu}$) from 30 MeV to 100 MeV\\
Unc\_Energy\_Flux30\_100 & E & erg cm$^{-2}$ s$^{-1}$ & 1$\sigma$ error on energy flux ($\nu F_{\nu}$) from 30 MeV to 100 MeV\\
Energy\_Flux100\_300 & E & erg cm$^{-2}$ s$^{-1}$ & Energy flux ($\nu F_{\nu}$) from 100 to 300 MeV \\
Unc\_Energy\_Flux100\_300 & E & erg cm$^{-2}$ s$^{-1}$ & 1$\sigma$ error on energy flux ($\nu F_{\nu}$) from 100 MeV to 300 MeV\\
Flux30\_100 & E & cm$^{-2}$ s$^{-1}$ & Photon flux from 30 MeV to 100 MeV\\
Unc\_Flux30\_100 & E & cm$^{-2}$ s$^{-1}$ & 1$\sigma$ error on photon flux from 30 MeV to 100 MeV\\
Flux100\_300 & E & cm$^{-2}$ s$^{-1}$ & Photon flux from 100 to 300 MeV \\
Unc\_Flux100\_300 & E & cm$^{-2}$ s$^{-1}$ & 1$\sigma$ error on photon flux from 100 MeV to 300 MeV\\
CLASS1 & 7A & --- &  Class designation for the associated source in the 3FGL catalog\\
Redshift & E & --- & Redshift for the associated source in the 3LAC catalog\\
%ASSOC1 & 18A & --- & Name of identified or likely associated sources in the 3FGL catalog\\
ASSOC\_3FGL & 18A & --- &Associated source in the 3FGL catalog\\
%ASSOC\_3LAC & 18A & --- & Associated source in the 3LAC catalog\\
ASSOC\_COMPTEL & 25A & --- &Associated source in the first COMPTEL catalog\\
\hline
\end{tabular}
\caption{\label{fits_file_table} Description of the entries in the 1FLE catalog fits file. The fluxes are obtained using PGWave. 
Energy\_Flux100\_300  and Unc\_Energy\_Flux\_100\_300, as well the Flux100\_300  and Unc\_Flux100\_300, are set to 0 
if the source is detected in the energy range between 30 -- 100 MeV, but not detected in the range 100 -- 300 MeV. The Redshift is set to 0 if the source has not associated source in the 3LAC catalog or there is no redshift information. The 1FLE catalog fits file is available only in electronic format as supplementary material.}
\end{table*}

\begin{table*}[h]
\small
\centering
\begin{tabular}{ccc}
\hline
\hline
Description &  Associated & \\
 & designator & Number\\
\hline
Pulsar & psr & 12 \\
Pulsar wind nebula & pwn & 2 \\
Supernova remnant	& snr & 2 \\
Supernova remnant / pulsar wind nebula & spp & 5 \\
High mass binary & hmb & 1\\
BL Lac type of blazar & bll & 31 \\
Flat spectrum radio quasar type of blazar & fsrq & 98 \\
Narrow-line seyfert 1 & nlsy1 & 1 \\
Radio galaxy & rdg & 3 \\
Steep spectrum radio quasar & ssrq & 1 \\
Normal galaxy (or part) & gal & 1 \\
Blazar candidate of uncertain type & bcu & 13 \\
Unclassified & '' & 17 \\
Unassociated & - & 11 \\
\hline
Total in the 1FLE & & 198\\
\end{tabular}
\caption{\small \label{class_source}
Source classes of the 1FLE sources determined using the 3FGL associations.}
\end{table*}

\begin{table*}[h]
\small
\centering
\begin{tabular}{ccccccccccc}
\hline
\hline
Source name & GLON & GLAT & Err\_pos & Signif. &  $\nu F_{\nu}$(30-100 MeV) & $\nu F_{\nu}$ (100-300 MeV)&Comment\\
  & (deg) & (deg) & (deg) & ($\sigma$) & $10^{-12}$ erg cm$^{-2}$ s$^{-1}$ & $10^{-12}$ erg cm$^{-2}$ s$^{-1}$&\\
\hline
1FLE J2206$+$7040& 110.02& 12.06&  0.25& 4.38 & 23.75 $\pm$ 7.16 & 0.0 $\pm$ 0.0& Diffuse\\
1FLE J0330$+$3304 & 157.42 & -18.94 & 0.25 & 9.87 & 23.56 $\pm$ 7.10 & 0.0 $\pm$ 0.0& 3FGL sources \\
1FLE J0422$+$5243&  151.75 & 2.07 & 0.25 & 7.00 & 22.73 $\pm$ 6.85 & 0.0 $\pm$ 0.0& Gal. plane\\
1FLE J0647-0345 & 215.89 & -2.48 & 0.25 & 7.75 & 17.71 $\pm$ 5.34 & 0.0 $\pm$ 0.0& Gal. plane\\
1FLE J0655-1106& 223.33& -4.08&  0.25 & 4.01& 14.93 $\pm$ 4.94 & 4.07 $\pm$ 1.63& Gal. plane \\
1FLE J0522$+$3734 &  170.17 & 0.68 & 0.25 & 5.00 & 13.66 $\pm$ 4.52 & 0.0 $\pm$ 0.0& Gal. plane\\
1FLE J0637-0110& 212.35& -3.72&  0.25 & 4.80& 10.88 $\pm$ 3.6 & 0.0 $\pm$ 0.0& Gal. plane\\
1FLE J1033$+$1601& 224.87& 56.14&  0.25 & 3.65& 10.30 $\pm$ 3.41 & 0.0 $\pm$ 0.0&$\sigma < 4$\\
1FLE J2158-5424& 339.89& -48.37&  0.25 & 3.99& 8.51 $\pm$ 2.82 & 0.0 $\pm$ 0.0&$\sigma < 4$\\
1FLE J1203-2504& 289.40& 36.53&  0.25 & 4.07& 8.39 $\pm$ 2.77 & 0.0 $\pm$ 0.0& 3FGL sources\\
1FLE J1030-3133& 270.81& 22.38&  0.25 & 3.43& 7.11 $\pm$ 2.35 & 0.0 $\pm$ 0.0& $\sigma < 4$\\
\hline
\end{tabular}
\caption{\small \label{new_source}
1FLE sources that do not have an association in the 3FGL catalog. 
For a more detailed description, see Section \ref{sec:unass_sources}.}
\end{table*}

%\clearpage
%\twocolumn

\subsection{Comparison with 3FGL}
In this section, we compare the 1FLE sources with the 3FGL catalog sources \citep{2015ApJS..218...23A}.
The 3FGL is based on the first four years of the \textit{Fermi}-LAT data between 100 MeV and 300 GeV using the Pass7 reprocessed event reconstruction. 
It contains 3033 sources: 33\% have no high-confidence counterparts at other wavelength and more than 1100 of the identified or associated sources are active galaxies of the blazar class.
%The classes that we use to categorize the 1FLE sources are listed in Table \ref{class_source} along with the numbers of sources assigned to each class. 
The 1FLE contains 198 sources and 187 of them have associations in the \Fermi-LAT 3FGL catalog.
Analysis with 1\fdg8 wavelet transform gives 174 sources:
144 sources are at high latitudes (four have no counterpart in 1\fdg4 analysis)  and 30 sources are at low latitudes (four have no counterpart in 1\fdg4 analysis).
Analysis with 1\fdg4 wavelet transform gives 168 sources:
132 sources are at high latitudes (five have no counterpart in 1\fdg8 analysis) 
and 36 sources are at low latitudes (three have no counterpart in 1\fdg8 analysis).
Combining the results of analysis with 1\fdg4 and 1\fdg8 wavelet scale, there are five sources at low latitudes and six sources at high latitudes without associations in the 3FGL catalog.
Among the associated sources, 148 are extragalactic, 22 are Galactic, and 17 are unclassified in the 3FGL.
More details about the 1FLE sources can be found in Tables \ref{class_source} and \ref{new_source}  and in Figure \ref{1fle_ps}.

For 94\% of the sources contained in the 1FLE we have found an association in the 3FGL catalog using our algorithm for the association  with a tolerance radius of 1\fdg5 
(for the association method see Section \ref{association_algorithm}).
The much smaller number of sources in 1FLE compared to, for example, the 3FGL catalog is due to several factors:
the effective area of the instrument between 30 -- 100 MeV ($<$0.35 m$^{2}$)  is much smaller than above 1 GeV ($>$0.9 m$^{2}$);
the angular resolution, even for the PSF3 event type, is larger than $3\adeg$ for energies below 100 MeV, 
while the angular resolution above 1 GeV is better than 0\fdg5;
in the analysis we use a wavelet filtering method rather than the maximum likelihood, 
which takes into account the precise shape of the point-spread function.

In spite of the different energy range and the longer time interval of 1FLE relative to 3FGL, all significant 1FLE sources have associations with 3FGL sources. 
One of the main reasons for that is the lower sensitivity of the \Fermi LAT at low energies due to smaller effective area and worse angular resolution compared to higher energies (the \Fermi-LAT PS sensitivity is the best around a few GeV, see Figure \ref{sensitivity_1fle}).  
We have also compared the 1FLE to the preliminary \Fermi-LAT eight year list of sources (FL8Y) \footnote{https://fermi.gsfc.nasa.gov/ssc/data/access/lat/fl8y/}.  
The FL8Y has a similar time interval as our analysis. There are several sources in FL8Y which have a flux extrapolated below 100 MeV above the statistical sensitivity threshold of 1FLE. 
Most of these sources are in the Galactic plane, where the wavelet transform has difficulty in separating the PS from the diffuse background. 
The sources above a galactic latitude of $|b| = 10^\circ$ are either relatively close to to the plane (within $|b| = 15^\circ$) or  close to other bright sources so that they fall in the negative tail of the wavelet transform of the bright source. 
There are two consequences of the non-observation of new sources in 1FLE: there were no sufficiently bright flaring sources after the 3FGL time interval, which were not already detected by 3FGL, and there are no bright sources with a very soft spectrum, for example, a cutoff around 100 MeV: sources of this type would not be detected in either the 3FGL catalog or the FL8Y list.

%Since all 1FLE sources are also detected in the 3FGL, this means that there are no soft and, at the same time, sufficiently bright PS to be detected in the 1FLE. In particular, there are no bright AGNs with a cut off around 100 MeV.

\subsubsection{1FLE Blazars}
\label{sec:redshift}
The 1FLE contains 148 extragalactic sources that are associated to the 3FGL ones.
The 3LAC catalog \citep{2015ApJ...810...14A} contains the 3FGL AGNs, located at high Galactic latitudes ($\mid b\mid > 10\adeg$), that have been detected in the energy range 100 MeV -- 300 GeV, between August 4, 2008 and July 31, 2012. The 3LAC includes 1591 objects with 467 (29\%) FSRQs, 632 (40\%) BL Lacs, 460 (29\%) BCUs and 32 (2\%) non-blazar AGNs.

The 1FLE blazars are subdivided into FSRQ and BL Lac in a different proportion with respect to the 3LAC ones. Among the 148 extragalactic sources contained in the 1FLE, 98 (66\%) are associated to FSRQ, 31 (21\%)  to BL Lacs, 13 (9\%) to BCUs and 6(4\%) to non-blazar AGNs.
The much larger fraction of FSRQs is to be expected, since we have studied the sources at lower energies than 3LAC. Consequently, our method is more sensitive to soft spectrum PS, which is typical for FSRQs compared to BL Lacs.

\begin{figure}[h]
\centering
\includegraphics[scale=\onepic]{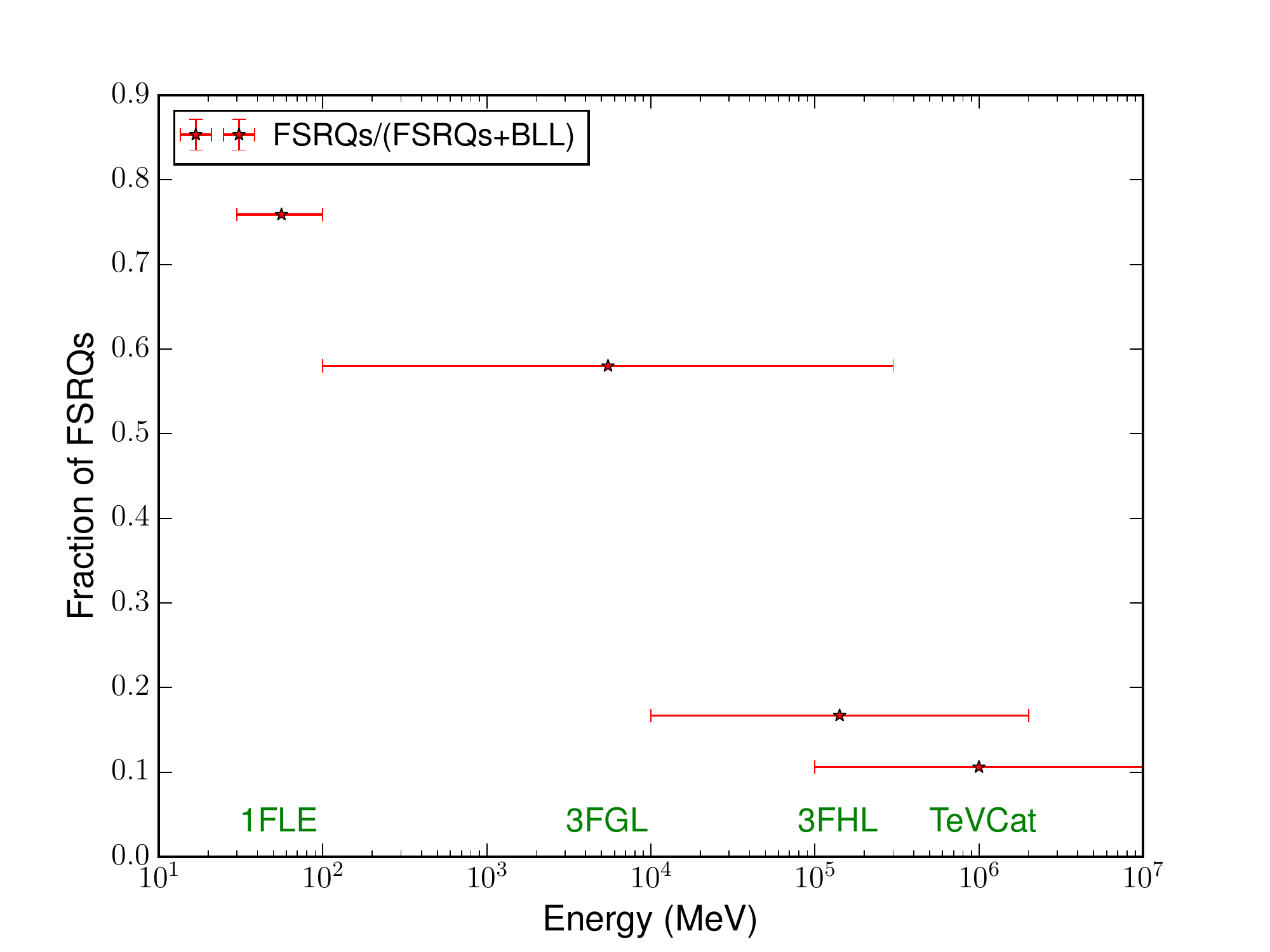}
\caption{\small \label{ratio_agn}
Fraction of FSRQs with respect to the sum of FSRQs and BL Lacs contained in each gamma-ray catalog versus the corresponding energy range used for the point source analysis.
}
\end{figure}

Figure \ref{ratio_agn} shows the fraction of FSRQs with respect to the sum of FSRQs and BL Lacs varying the energy range used for the point source analysis in the following gamma-ray catalogs: 1FLE, 3LAC, 3FHL \citep{2017ApJS..232...18A} and TeVCat%
\footnote{http://tevcat.uchicago.edu/}.
The ratio of BL Lac (FSRQ) blazars increases (decreases) with the observed energy.

\begin{figure}[h]
\centering
\includegraphics[scale=\onepic]{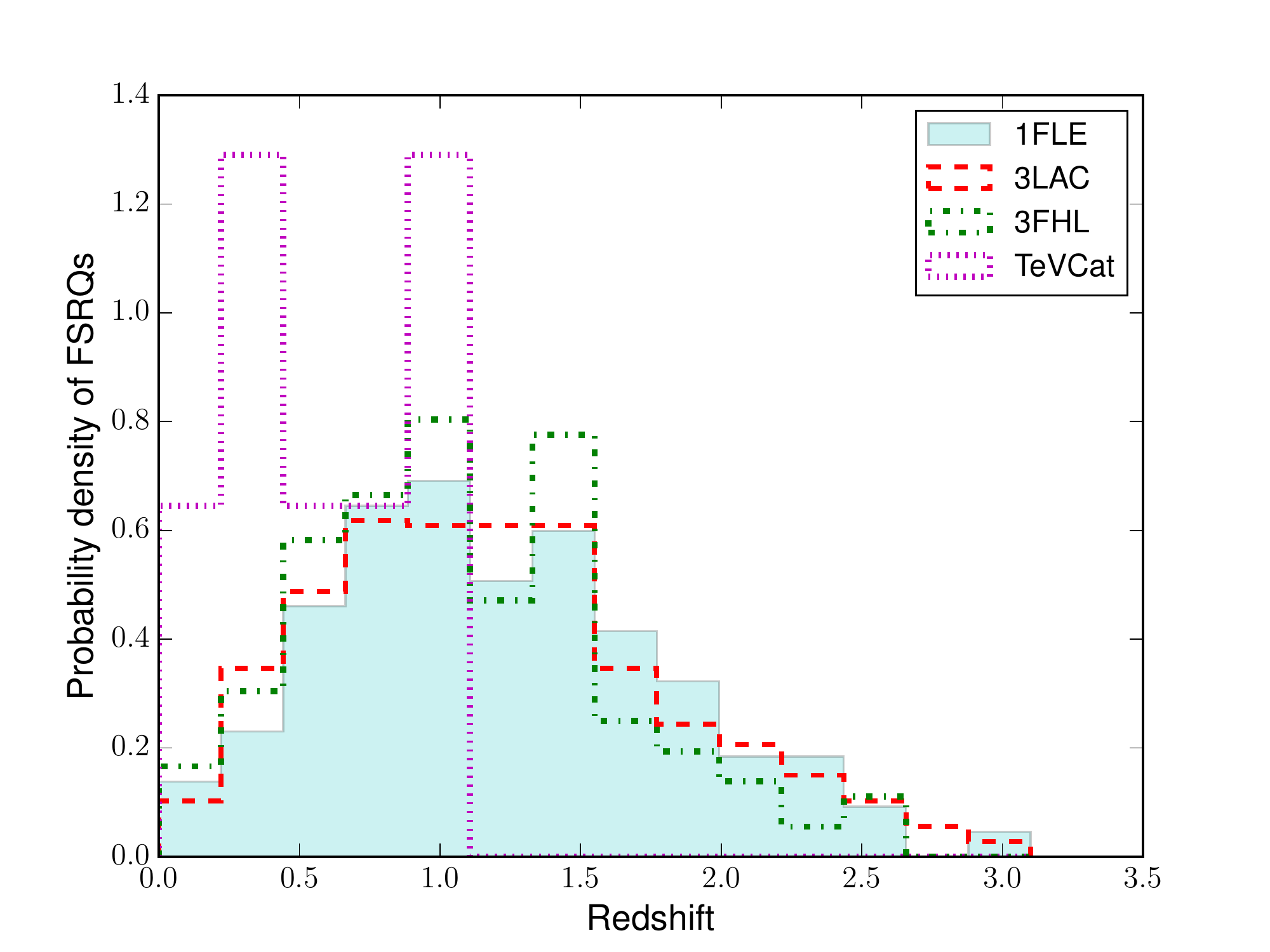}
\caption{\small \label{redshift_fsrq}
Distributions of known redshifts for FSRQs.
}
\end{figure}

\begin{figure}[h]
\centering
\includegraphics[scale=\onepic]{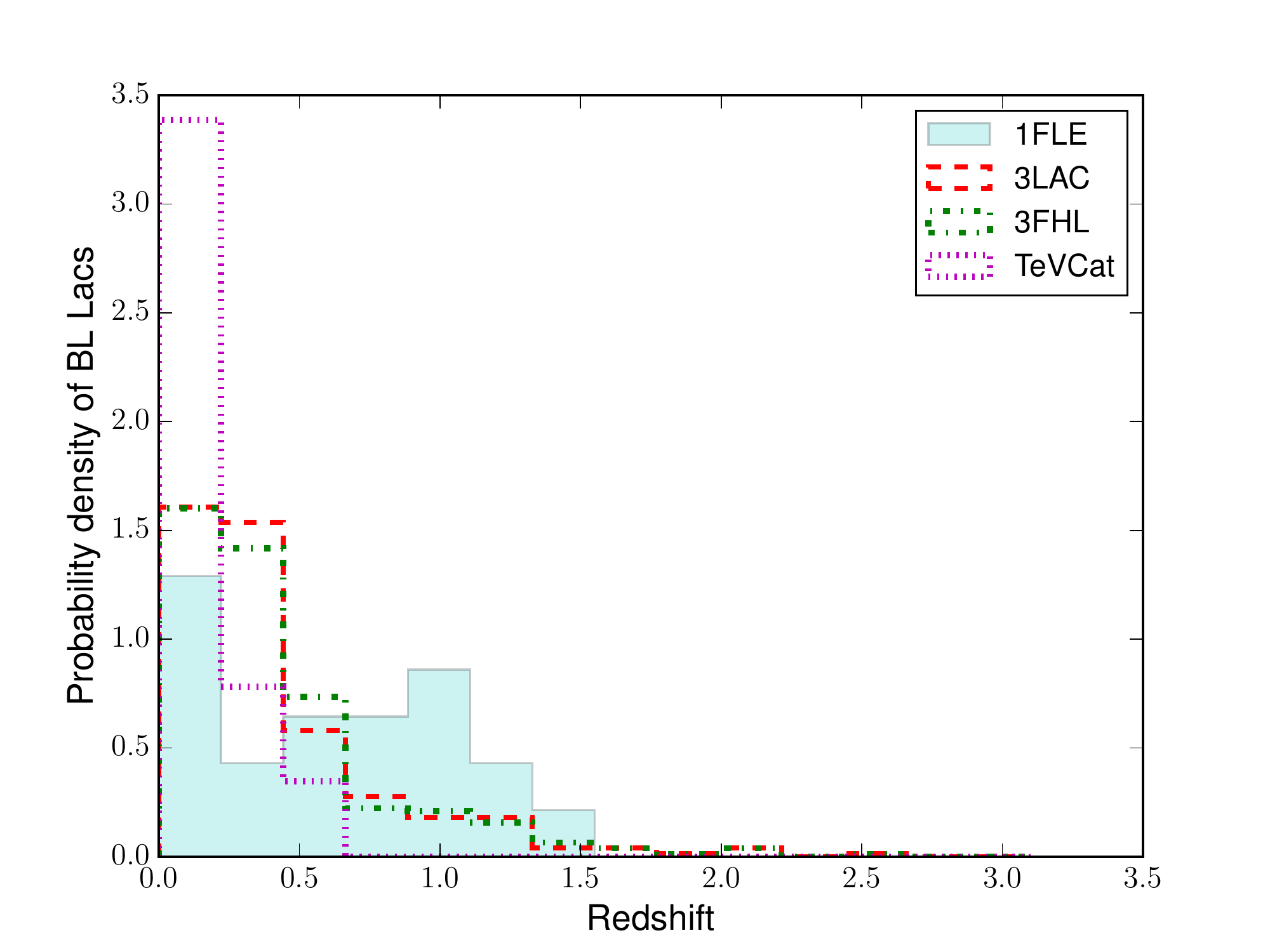}
\caption{\small \label{redshift_bll}
Distributions of known redshifts for BL Lacs.
}
\end{figure}

Table \ref{table_redshift} shows the mean values of the redshift for each class of blazars.
We used a Kolmogorov-Smirnov (KS) test to compare the distributions of redshifts for each
type of blazar in 3LAC compared to the subset of blazars of this type that is also found in 1FLE.
Considering the redshift distribution of all blazars, the two distributions (1FLE and 3LAC) have different averaged redshift ($z_{1FLE} =1.06 \pm 0.06$ and   $z_{3LAC} =0.84 \pm 0.02$) and they are not consistent with each other according to
the KS test, which has the p-value of $8.1 \times 10^{-6}$. 
The distributions of the FSRQ are consistent between the two catalogs (p-value of the KS test is 0.964),
while the BL Lac distributions are not fully consistent: p-value of the KS test is 0.018 (see also Figures \ref{redshift_fsrq} and \ref{redshift_bll}). 
The redshift distribution of FSQR in the 1FLE is consistent with the FSRQ redshift distributions in the other catalogs (apart from TeVCat, due to the small number of FSRQs in the catalog).
The 1FLE BL Lacs have an average redshift ($z_{1FLE} =0.59 \pm 0.09$),
which is larger than the average redshift of all BL Lacs in the 3LAC ($z_{3LAC}=0.41 \pm 0.02$).
This is in agreement with the larger number of low-synchrotron peaked blazar type in the 1FLE (18 out of 31), which have softer spectra and higher redshifts than BL Lacs on average, with respect to the same one in the 3LAC (162 out of 632).
The average redshift for the BL Lacs (Figure \ref{redshift_bll}) is larger than the average redshift for these sources
in the other catalogs.

\begin{table*}[h]
\small
\centering
\begin{tabular}{cccc}
\hline
\hline
Blazar class &  1FLE  & 3LAC & KS test\\
  &  z$_{av}$ &  z$_{av}$ & p-value\\
\hline
All blazars &  1.06 $\pm$ 0.06 &  0.84 $\pm$ 0.02 & $8.1 \times 10^{-6}$\\
FSRQ  & 1.22 $\pm$ 0.06  & 1.21 $\pm$0.03 & 0.964\\
BL Lac & 0.59 $\pm$ 0.09 & 0.41 $\pm$ 0.02  & 0.018\\
Other blazars & 0.55 $\pm$ 0.17 & 0.33 $\pm$0.04 & 0.124\\
\hline
\end{tabular}
\caption{\small \label{table_redshift}
1FLE Blazar classes and corresponding average redshift determined using the 3LAC association. 
The label 'KS test' refers to the value of the Kolmogorov-Smirnov test for the redshift distributions of the 
1FLE and 3LAC sources of the considered class of blazars.}
\end{table*}

\subsubsection{1FLE sources not associated to the 3FGL}
\label{sec:unass_sources}
Table \ref{new_source} reports the information about the 11 1FLE sources that have no association in the 3FGL catalog.
Among those sources, five (1FLE J0422$+$5243, 1FLE J0647-0345, 1FLE J0655-1106, 1FLE J0522$+$3734 and 1FLE J0637-0110) are inside the Galactic plane ($\mid b\mid <10\adeg$) where the diffuse model has various structures that could influence the background elimination in the PGWave tool and six of them are outside the Galactic plane.

The 1FLE J2206$+$7040 source is inside a particular region where the diffuse emission has some bright features; however there are no evident peaks of WT of the diffuse map in that position. The 1FLE J0330$+$3304 and 1FLE J1203-2504 sources are surrounded by 3FGL sources. 
Although they do not have an association in the 3FGL, they are probably connected to the 3FGL sources: due to the large PSF in the energy range 30 -- 100 MeV, if there are two or more sources close to each other, they could form a single structure in the counts map and PGWave does not distinguish the different sources but returns a seed in the middle. There are no 3FGL sources in the local region around 1FLE J1033$+$1601, 1FLE J2158-5424 and 1FLE J1030-3133. They are therefore good candidates for new sources in the gamma energy band, although the significance is less than 4$\sigma$.

\subsection{Comparison with 1st COMPTEL Catalog}
COMPTEL provided the first complete all-sky survey in the energy range 0.75 --  30 MeV. 
The first COMPTEL catalog \citep{2000A&AS..143..145S} %is restricted to published results. 
contains 26 steady sources: nine pulsars, four other Galactic sources, ten AGNs and three unidentified high-latitude sources.  
Eight out of the ten COMPTEL AGNs are associated to the 1FLE sources. 

Regarding the two COMPTEL AGNs without association in the 1FLE, one (PKS 0528+134) is close to the Galactic plane in a region of significant structures in the diffuse emission and the other one (GRO J0516-609) is a flaring source observed by COMPTEL only between 1--10 MeV, it has no association also in the 3FGL.
The 1FLE contains five pulsars detected also by COMPTEL: Crab, Geminga, Vela, PSR B1055-52 and GRO J2227+61.
Most of the COMPTEL sources without an association in the 1FLE are in the Galactic plane, so they could be masked, in our analysis, by peaks of the diffuse. 
Among the other Galactic sources contained in the COMPTEL catalog, Nova Per 1992, which is a transient X-ray binary, and Cygnus X-1, which is a persistent X-ray binary, have no associations also in the 3FGL catalog.
All the COMPTEL unidentified sources have association neither in the 3FGL nor in any other \Fermi source catalog.

\begin{figure}[h]
\centering
\includegraphics[scale=\onepic]{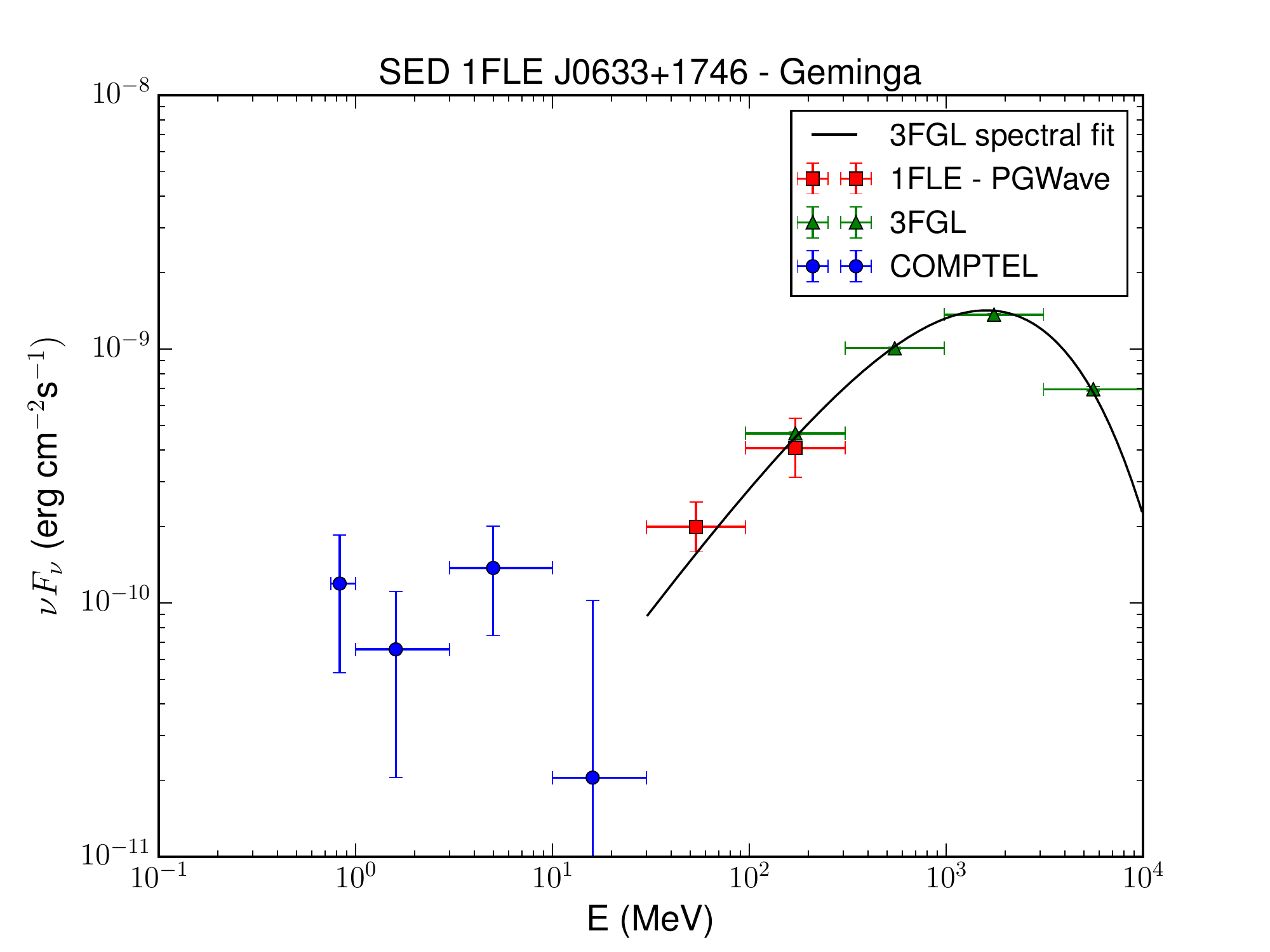}
\caption{\small \label{sed_geminga}
Spectral energy distribution of Geminga. 
Red squares: flux derived with the PGWave analysis (interval time 2008 - 2017), 
blue circles: COMPTEL flux values (interval time 1991 - 2000),
green triangles: flux from the 3FGL catalog (interval time 2008 - 2012), black line: 3FGL spectral fit.
}
\end{figure}

\begin{figure}[h]
\centering
\includegraphics[scale=\onepic]{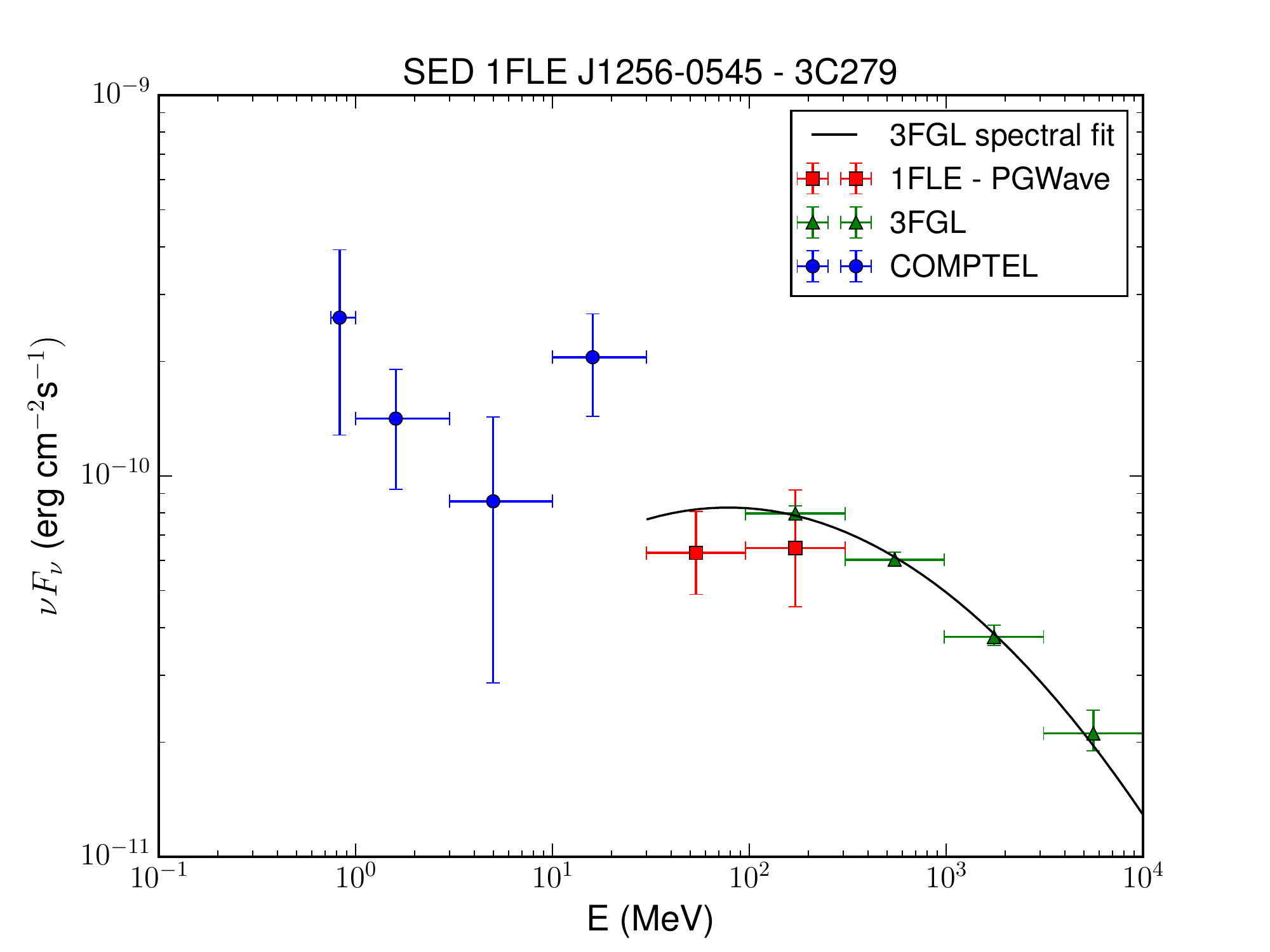}
\caption{\small \label{sed_3c279} Spectral energy distribution of 3C279. 
The labels are the same as in Figure \ref{sed_geminga}.
}
\end{figure}

Figures \ref{sed_geminga} and \ref{sed_3c279} show examples of the spectral energy distributions for some of the gamma-ray sources detected in 1FLE. 
We also report, for a comparison, fluxes from the COMPTEL and the 3FGL catalogs.

\subsection{Comparison of 1FLE fluxes with 3FGL}
\label{sec:1fle_sed}
\begin{figure}[h]
\centering
\includegraphics[scale=\onepic]{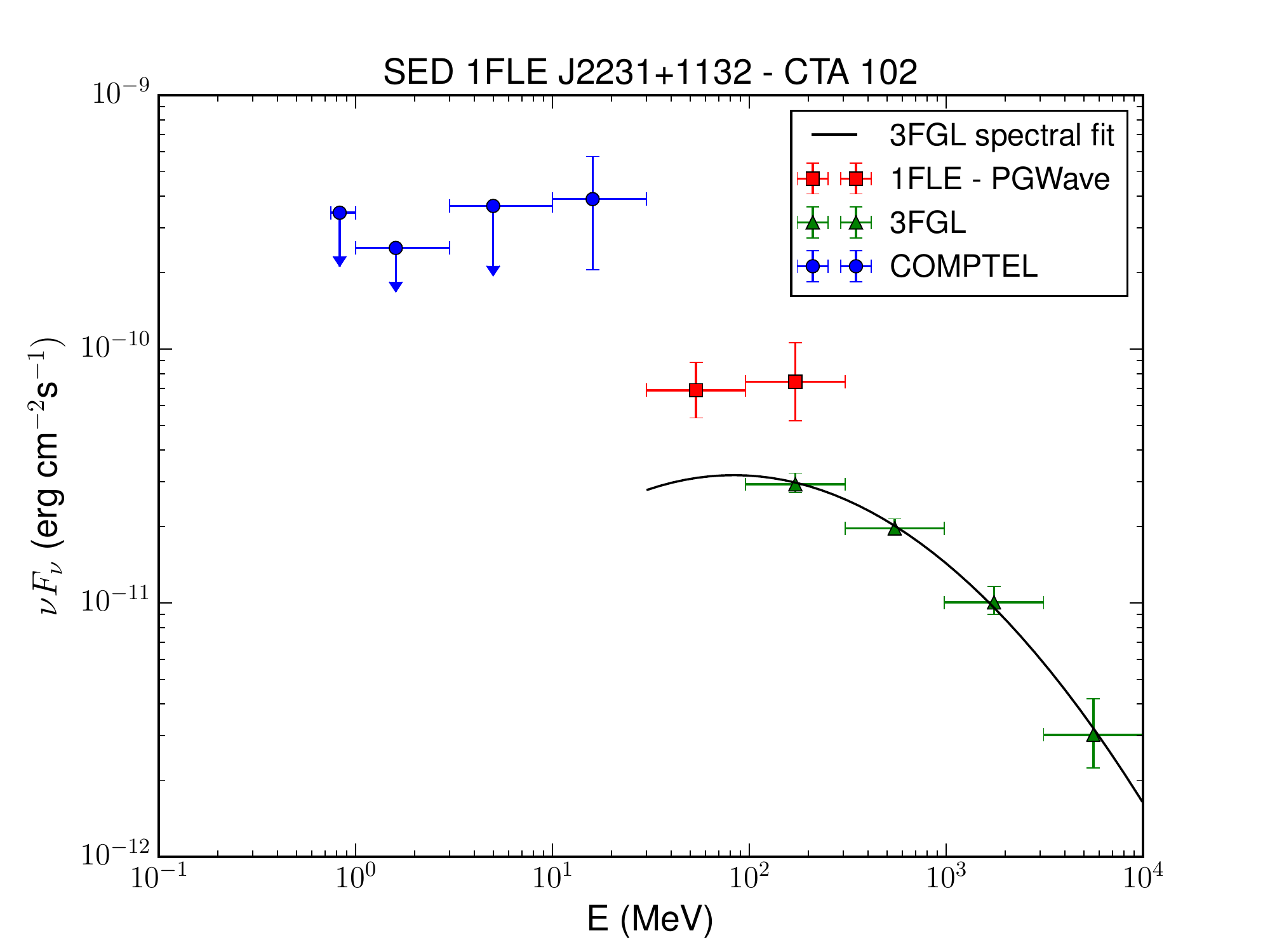}
\caption{\small \label{sed_3cta102} Spectral energy distribution of 1FLE J2231+1132 (CTA 102). 
The labels are the same as in Figure \ref{sed_geminga}. 
The 1FLE J2231+1132 source, also called CTA 102, had a flare after the 3FGL observation time and the 1FLE flux points are consistent with the estimated flux in the FL8Y list.
}
\end{figure}

There are several sources in the 1FLE catalog whose fluxes in the 30 - 100 MeV and 100 - 300 MeV bands differ significantly from the fluxes of the associated 3FGL sources. 
Several of these sources are associated with AGNs, which had flares either during the 3FGL observation time
or after the 3FGL observation time (see Table \ref{flare_table} and e.g., Figure \ref{sed_3cta102}).
%\onecolumn
\begin{table*}[h]
\small
\centering
\begin{tabular}{cccccc}
\hline
\hline
Source Name & GLON & GLAT & 1FLE $\nu F_{\nu}$(100-100 MeV) & 3FGL $\nu F_{\nu}$ (100-300 MeV) & Flare comment\\
  & (deg) & (deg) & $10^{-12}$ erg cm$^{-2}$ s$^{-1}$ & $10^{-12}$ erg cm$^{-2}$ s$^{-1}$ & \\
\hline
1FLE J0424-0042 & 194.8 & -32.6 & 5.49 $\pm$ 2.19  & 18.79 $\pm$ 1.17 & flare in 3FGL\\
1FLE J0443-0024 & 197.5 & -28.2 & 6.26 $\pm$ 2.50 & 19.72 $\pm$ 0.86 & flare in 3FGL\\
%1FLE J0910+0159 & 228.5 & 31.5 &  flare in 3FGL\\
1FLE J1224+2118 & 255.5 & 81.6 & 49.77 $\pm$ 14.79 & 83.52 $\pm$ 1.12 & flare in 3FGL\\
1FLE J1227+0218 & 289.1 & 64.6 & 37.46 $\pm$ 10.63 & 87.53 $\pm$ 1.41 &  flare in 3FGL\\
1FLE J1332-0518 & 321.6 & 56.0 & 11.93 $\pm$ 3.39 & 26.22 $\pm$ 1.93 & flare in 3FGL\\
1FLE J1503+1033 & 11.3  & 54.8 & 3.60 $\pm$ 1.02 & 6.73 $\pm$ 1.19 & flare in 3FGL\\
1FLE J2231+1132 & 77.1 & -38.6 & 74.32 $\pm$ 22.09 &  29.34 $\pm$ 1.03 & flare after 3FGL\\
\hline
\end{tabular}
\caption{\small \label{flare_table}
1FLE sources with a flare during the 3FGL observation time (flare in 3FGL) or after the 3FGL observation time (flare after 3FGL).}
\end{table*}
%\clearpage
%\twocolumn

For these sources, we get flux estimates closer to the fluxes given in FL8Y.

For other sources the difference can be attributed either to the presence of several 3FGL sources close to each other and unresolved by
the wavelet transform (i.e., 1FLE J0329-3724, 1FLE J1047+7131 and 1FLE J1838+6812) or by the presence of a bright source within 7 - 15 degrees.
The negative tail of the wavelet kernel from the bright source reduces the value of the WT peaks for the nearby sources,
which leads to an underestimate of the corresponding fluxes. This is the case of 1FLE J0531+0707, which is in the vicinity of the Crab pulsar, and 1FLE J2254+1617, which has a bright source 1FLE J2231+1132 within 8$\adeg$.

\section{Conclusions}
We have analyzed the first 8.7 years of \textit{Fermi}-LAT Pass 8 data and derived, for the first time, a catalog of sources in the energy range  30 -- 100 MeV.
The 1FLE catalog, described in this paper, contains 198 sources detected using PGWave, a background-independent wavelet-based method.
%Comparison
This catalog closes the gap between the previous gamma-ray catalogs: 
the COMPTEL observations at energies lower than 30 MeV and the \Fermi-LAT catalogs at energies higher than 100 MeV.
For 94\% of the 1FLE sources we have found an association in the 3FGL \Fermi-LAT catalog.
Among the 11 sources without associations in the 3FGL, five are within $|b| < 10\adeg$ where the Galactic diffuse emission has several structures. Considering the six sources at high latitude: one source is in a region with a large Galactic emission,
two sources are surrounded by 3FGL sources (which are likely to merge in a single source due to large PSF at low energies),
and three sources have a significance between 3.5 and 4 $\sigma$.
% the remaining sources are compatible with the false positive rate (2\%).
The ratio of FSRQs to BL LACs varies from approximately three to one in the 1FLE, to 1 to 1 in the 3FGL and one to six in the 3FHL.
The redshift distribution of the BL Lacs in the 1FLE is peaked toward higher redshift with respect to the same one for the 3FGL BL Lacs.
This is correlated to the large ratio of low-synchrotron peaked BL Lacs in the 1FLE (58\%) with respect to the same in the 3LAC  (26\%). They have in fact softer spectra and higher redshifts than BL Lacs on average.

\begin{figure}[h]
\centering
\hspace{-3mm}
\includegraphics[scale=\onepic]{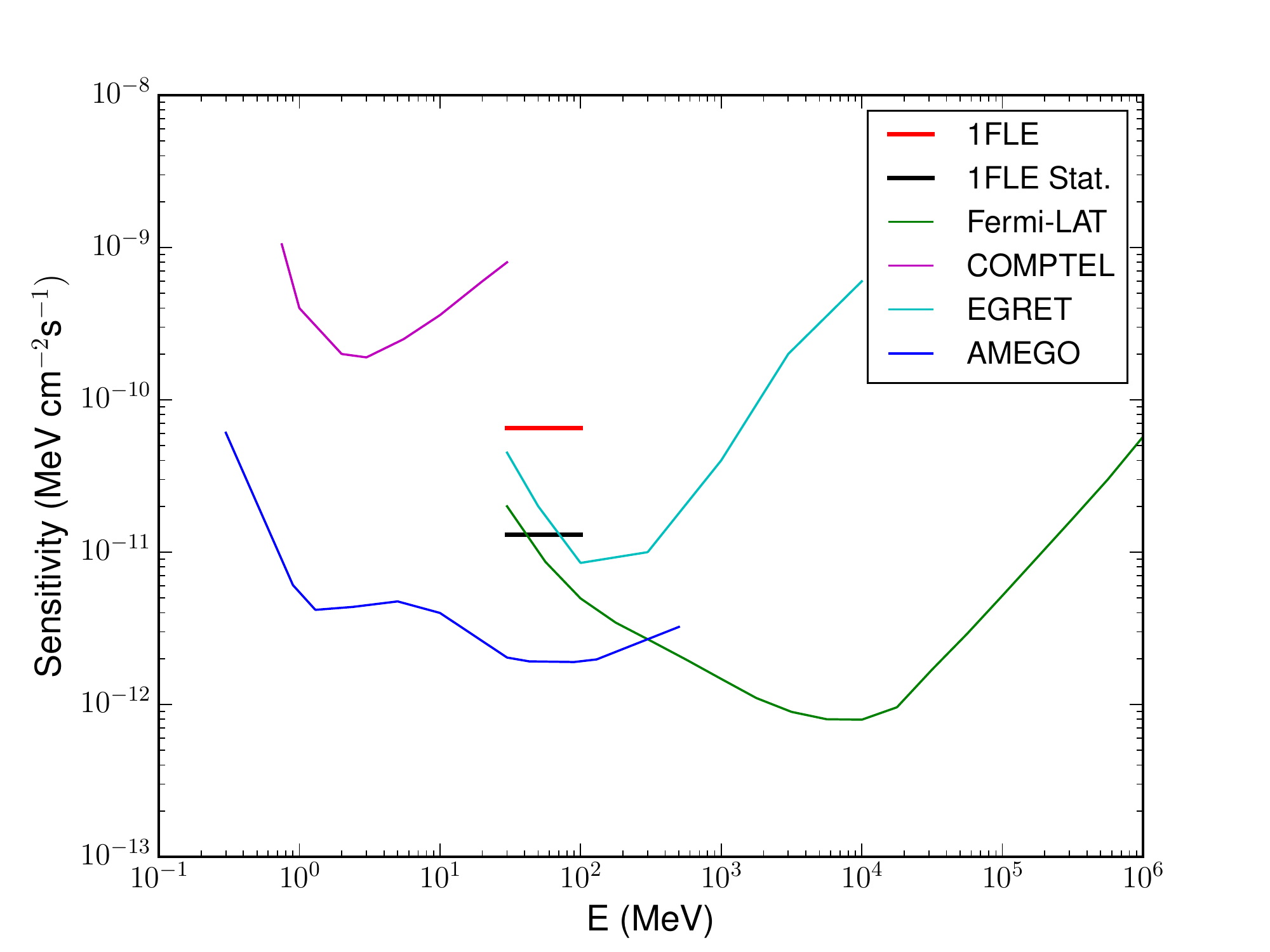}
\caption{\small \label{sensitivity_1fle}
Comparison of the PS sensitivity of the 1FLE catalog and the differential sensitivities of different gamma-ray instruments. 
The COMPTEL (magenta line) and EGRET (cyan line) sensitivities are given for the typical observation time accumulated during the nine years of the \textit{CGRO} mission. The \Fermi-LAT sensitivity (green line) is for a high Galactic latitude source in ten years of observation in survey mode. The blue line represents the simulated continuum sensitivity (3$\sigma$, 3 years) for AMEGO. 
In red, the 1FLE total sensitivity (see also Figure \ref{detection_flux}), while the black represents the 1FLE statistical sensitivity determined as the flux corresponding to the 5$\sigma$ significance of PGWave (Figure \ref{significance}).}
\end{figure}

% refer to eASTROGAM
An instrument with better angular resolution, such as AMEGO \citep{2017HEAD...1610313M} or e-ASTROGAM \citep{2017ExA...tmp...24D, 2017arXiv171101265D}, will improve markedly the sensitivity 
at energies below 100 MeV and increase the number of sources detected at MeV energies.
%\footnote{See \small https://pcos.gsfc.nasa.gov/physpag/probe/AMEGO\_probe.pdf}
In Figure \ref{sensitivity_1fle}, we compare the sensitivity of the 1FLE catalog to the PS sensitivities of various gamma-ray experiments.
At energies below 100 MeV, both AMEGO and e-ASTROGAM are expected to have a sensitivity which is more than two times better than that of \Fermi-LAT.
These new instruments are able to widely extend our knowledge of the MeV sky.

% Acknowledgement
\subsection*{ACKNOWLEDGMENTS}
The effort of the LAT-team Calibration \&  Analysis Working Group to develop Pass 8, and also the LAT Catalog Working Group for the support given to this project, are gratefully acknowledged.

The Fermi LAT Collaboration acknowledges generous ongoing support from a number of agencies and institutes that have supported both the development and the operation of the LAT as well as scientific data analysis. These include the National Aeronautics and Space Administration and the Department of Energy in the United States, the Commissariat à l'Energie Atomique and the Centre National de la Recherche Scientifique / Institut National de Physique Nucléaire et de Physique des Particules in France, the Agenzia Spaziale Italiana and the Istituto Nazionale di Fisica Nucleare in Italy, the Ministry of Education, Culture, Sports, Science and Technology (MEXT), High Energy Accelerator Research Organization (KEK) and Japan Aerospace Exploration Agency (JAXA) in Japan, and the K. A. Wallenberg Foundation, the Swedish Research Council and the Swedish National Space Board in Sweden.

Additional support for science analysis during the operations phase is gratefully acknowledged from the Istituto Nazionale di Astrofisica in Italy and the Centre National d'Etudes Spatiales in France. This work performed in part under DOE Contract DE- AC02-76SF00515.

\newpage
\bibliography{lowe_cat_papers}  

%\begin{appendix}
\appendix
\section{Monte Carlo simulations}

\subsection{PGWave: a wavelet transform method}
\label{pgwave}

PGWave is an image-based source detection technique. 
It is based on the wavelet transform function \citep{1997ApJ...483..350D}. 
PGWave uses the 2-dim Mexican Hat wavelet. This ensures that the Wavelet Transform (WT) of a function $f(x,y)=c_{1}+c_{2}x+c_{3}y$ (a tilted plane) is zero. Therefore the WT will be zero for both a constant or uniform gradient local background. 

The peak of the WT for a source with a Gaussian shape ($N_{src}$ total counts and width $\sigma_{src}$) is
\begin{equation}
w_{peak}(a) = \dfrac{2 N_{src}}{\left( 1 + \dfrac{\sigma_{src}^{2}}{a^{2}}\right)^{2}},
\end{equation}
where $a$ is the scale of the wavelet transform, also called WT scale.
There is hence a linear correlation between the WT peak value $w_{peak}$ and the total number of photons detected from a source $N_{src}$.
Results of preliminary studies on using the PGWave to estimate the flux of gamma-ray PS without the need of a background model
were reported in \cite{2016arXiv161001351P}.

\subsection{Description of Monte Carlo simulations}
\label{appendix_mc_simulations}
%General setup of the simulations.

The large PSF below 100 MeV implies that the number of independent positions in the sky, namely the positions that we could spatially distinguish with that resolution, is not very large. 
Therefore, it is necessary to perform an accurate study on the choice of the event selection (PSF event types) and on the parameters of the point-source analysis (PGWave parameter) in order to optimize the detection and minimize the number of false positives.
We use Monte Carlo (MC) simulations to optimize the parameters of the analysis. 

The simulations have two steps: choice of the diffuse model and choice of the positions and fluxes of point sources.
For the diffuse model, we fit the gamma-ray data with a combination of templates that trace different components of emission,
such as hadronic interactions of cosmic rays with interstellar gas, inverse Compton (IC) scattering, \Fermi bubbles etc.
The construction of the diffuse model follows the same steps as the Sample model in \cite{2017ApJ...840...43A}.
In particular, the diffuse model has the following templates: $\pi^0$ + bremsstrahlung in 5 Galactocentric rings,
3 IC components corresponding to IR, starlight and CMB radiation fields, \Fermi bubbles at $|b| > 10\adeg$,
geometric Loop I template, Sun and Moon templates, and the isotropic template. 
In the fit to the data, we also add a point sources template derived  from the 3FGL catalog.
The only differences from the Sample model of \cite{2017ApJ...840...43A} are: 
we do not use a PS mask and the isotropic emission is forced to have a power-law spectrum (the index of the spectrum is fit to the data).
We fit the data between 31.2 MeV and 312 MeV in 6 logarithmic bins.
To derive the diffuse emission in two large energy bins 31.2 MeV -- 100 MeV and 100 MeV -- 312 MeV, 
we add the count rates of the diffuse components (all templates except the PS template) in the 3 bins below 100 MeV
and 3 bins above 100 MeV.

\vspace{5mm}
\begin{table}[h]
\centering
\begin{tabular}{|c|c|}
\hline
Parameter & Value\\
\hline
\hline
IRFs & P8R2\_SOURCE\_v6\\
\hline
Energy range & 30-100 MeV/ 100-300 MeV  \\
\hline
Pixel dimension & 0\fdg458\\
\hline
Interval of time & 8.7 years\\
\hline
Number of sources & 369\\
\hline
Flux & $10^{-8} - 10^{-4.5}{\rm \; cm^{-2}s^{-1}}$\\
\hline
\end{tabular}
\caption{ \label{simulation_map_par} Parameters used for the generation of MC maps.}
\end{table}

\begin{figure}[h]
\centering
\includegraphics[scale=\mapscale]{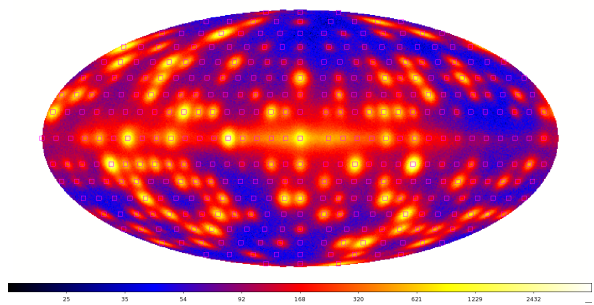}
\caption{\small \label{grid_map}
Count map of the first setup (flat $S dN/dS$, positions on a grid) which contain 369 PS in a grid with 10$\adeg$ separation.
}
\end{figure}

For the generation of point sources, we use two different setups. 
Both setups contain a population of 369 PS with a flux between 30 and 100 MeV randomly chosen from 
a flat distribution on the logarithmic scale between 
$10^{-8}{\rm \; cm^{-2}s^{-1}}$ (which is close to the threshold of the detection) and $10^{-4.5}{\rm \; cm^{-2}s^{-1}}$.
The parametric representation of the PS flux is obtained by taking the parametric representation of a random source in the 3FGL catalog
and rescaling it so that the integrated flux between 30 and 100 MeV is equal to the randomly chosen flux from the flat distribution.
In the first setup the PS are positioned in the sky in a grid with a separation of $10\adeg$ (Figure \ref{grid_map}), while in the second one they are randomly positioned in the sky (Figure \ref{random_map}).
Figure \ref{grid_map} and all the following figures refer to the energy bin 30 –- 100 MeV.
Table \ref{simulation_map_par} contains the main parameters that are used in the simulation.
The flux sampling from a flat distribution in the log space is useful for obtaining a good statistic of flux and position reconstruction
at high fluxes (see, e.g., Figure \ref{flux_reconstrucation_30}).
However, the corresponding source count distribution $dN / d\log S \propto const.$ is not realistic and may give detection and false positive rates that are not adequate for the expected populations of PS.

We verify, for this reason, the detection rate and the false positive rate expectations by including all 3FGL sources in the simulations,
where we use a random position in the sky for each source.The
fluxes are obtained by integrating the parametric representations in 3FGL. Similar rates are observed also in the simulated maps with all the 3FGL sources.

\begin{figure}[h]
\centering
\includegraphics[scale=\mapscale]{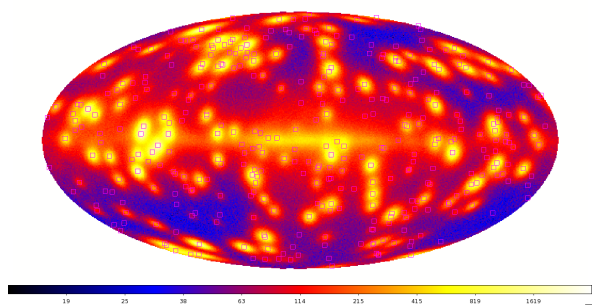}
\caption{\small \label{random_map}
Count map of the second setup (flat $S dN/dS$, random positions) which contains 369 PS randomly positioned in the sky.}
\end{figure}

\subsection{Selection of event type}
\label{sec:PSF_selection}

%Here we describe the Selection of the PSF classes.
At low energies, \Fermi LAT has a PSF that increases from $\sim 5\adeg$ at 100 MeV to $\sim 12\adeg$ at 30 MeV.
One can improve the resolution by selecting events with a better angular resolution (e.g., PSF3 event type),
but this subselection of events leads to a decrease of statistics.

In order to find the optimal combination of PSF event types, 
we compare the detection efficiencies and false positive rates for the following selections:
\bi
\item
all PSF event types combined together;
\item
PSF1, PSF2, and PSF3 event types; 
\item
PSF2 and PSF3 event types;
\item
only PSF3 event type.
\ei

We use the same analysis pipeline that is described in Section \ref{analysis_procedure}. 
One of the most important parameters is the wavelet transform scale.

\begin{figure}[h]
\centering
\includegraphics[scale=\onepicabs]{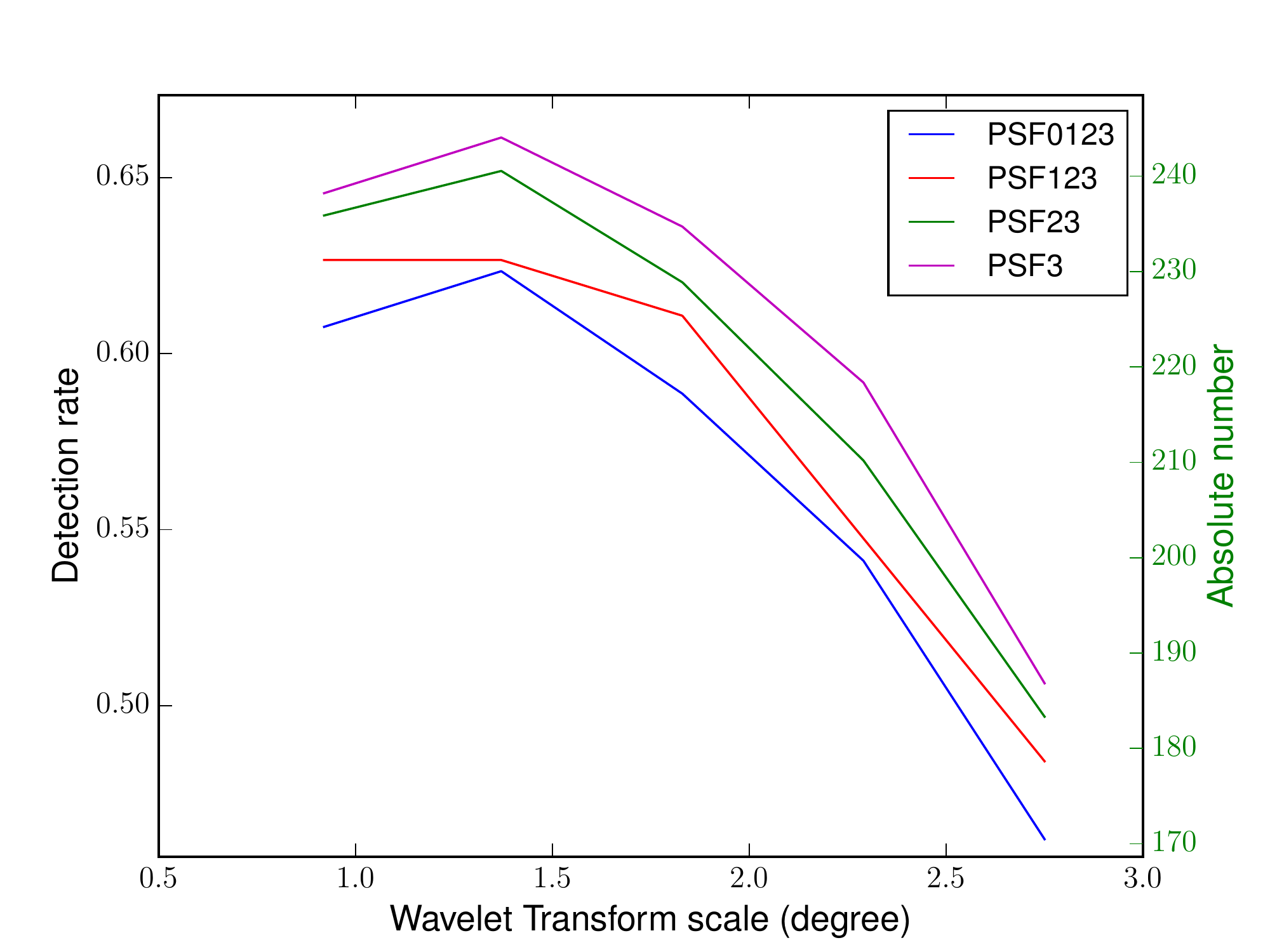}
\caption{\small \label{grid_detection}
Detection rate varying the dimension of the wavelet transform scale (setup containing 369 PS with flat $S dN/dS$ and random positions). The plot refers to the 30 -- 100 MeV band.}
\end{figure} 

\begin{figure}[h]
\centering
\includegraphics[scale=\onepicabs]{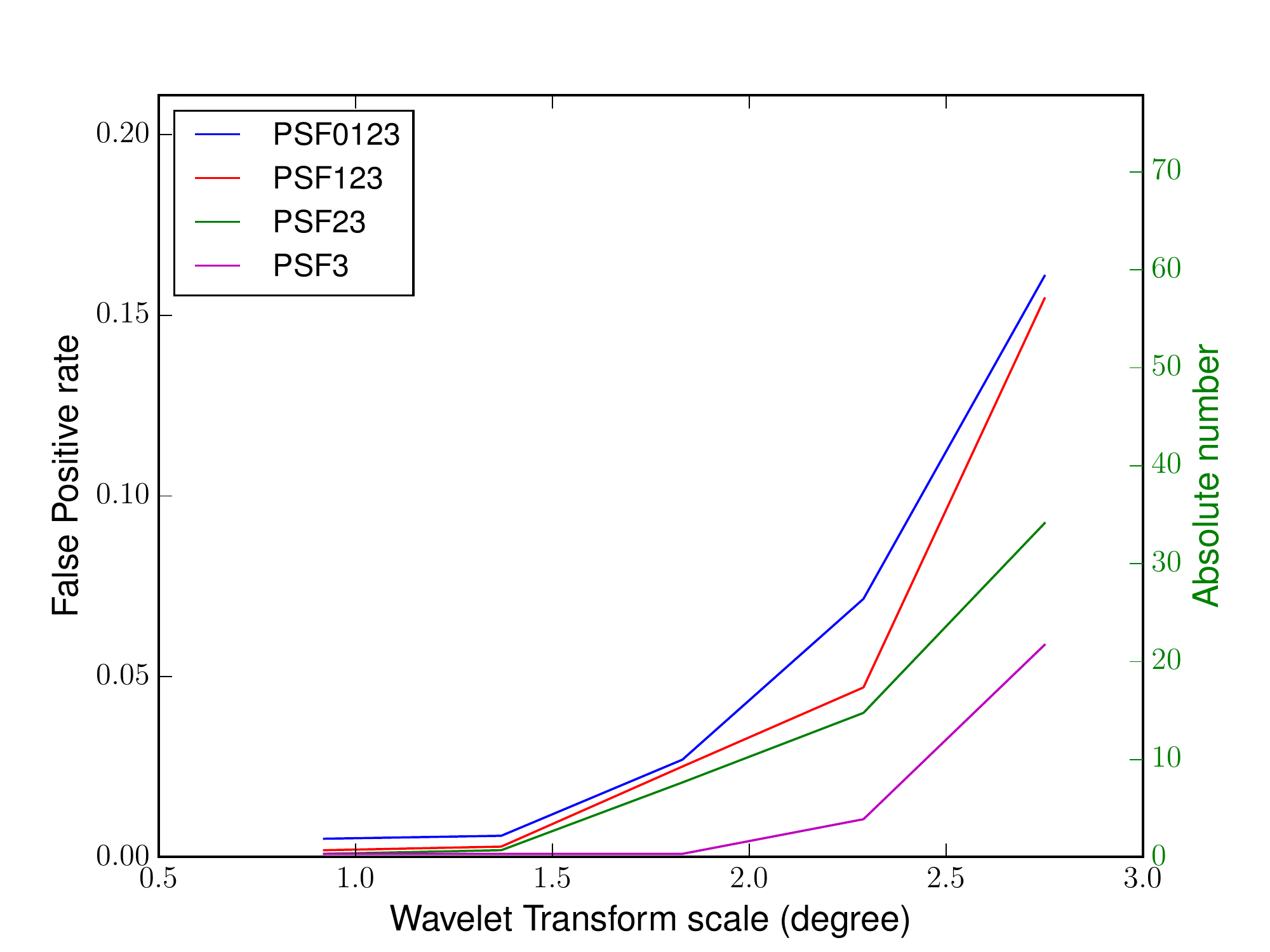}
\caption{\small \label{grid_false_pos}
False positives varying the dimension of the wavelet transform scale (setup containing 369 PS with flat $S dN/dS$ and random positions). The plot refers to the 30 -- 100 MeV band.}
\end{figure} 

Figures \ref{grid_detection} and \ref{grid_false_pos} contain the results of the analysis, using the grid setup, for different combinations of PSF event types, where we vary the WT scale. They show respectively the detection rate and the false positive rate as a function of the WT scale.
The plots show that for the PSF3 event type the detection rate is the largest while the false positive rate is the smallest. A similar behavior is observed in the analysis using the grid setup. 
Both the grid and the random positions setups show that the PSF3 event type, even if it has smaller statistics, gives the best detection and false positive rates.

\subsection{Selection of PGWave parameters}
\label{sec:PGWparam_selection}

In order to maximize the detection rate and minimize the false positive rate, we optimize the main parameters of our analysis.
For this analysis we use the simulated maps with 369 PS with flat $S dN/dS$ and random positions in the sky (see Section \ref{appendix_mc_simulations}).
The parameters that we optimize are reported in Table \ref{pgw_par}. The threshold was set at 3 $\sigma$. 
We perform the analysis separately for both energy bins: 30 -- 100 MeV and 100 -- 300 MeV.

We study the behavior of detection rate and false positive rate varying the wavelet transform scale, 
the minimum number of connected pixels to define a peak and the minimum distance between the sources. 
We vary also the merging radius, that is the tolerance radius for merging the seeds from different ROIs (see analysis description).

%\vspace{5mm}
\begin{table*}[h]
\centering
\begin{tabular}{|c|c|c|}
\hline
PGWave parameter & Values & Step\\
\hline
\hline
MH wavelet transform scale & 0\fdg9 - 3\fdg6 & 0\fdg46  \\
\hline
Min number of connected pixels & $2 - 8$ & 1 \\
\hline
Min distance between sources & 1\fdg8 - 2\fdg7 & 0\fdg46\\
\hline
\end{tabular}
\caption{ \label{pgw_par} List of PGWave parameters that we use to optimize our analysis. 
Considering also the variation of the merging radius, we try more than 170 different combinations of the analysis parameters.}
\end{table*}

\begin{figure}[h]
\centering
\includegraphics[scale=\onepicabs]{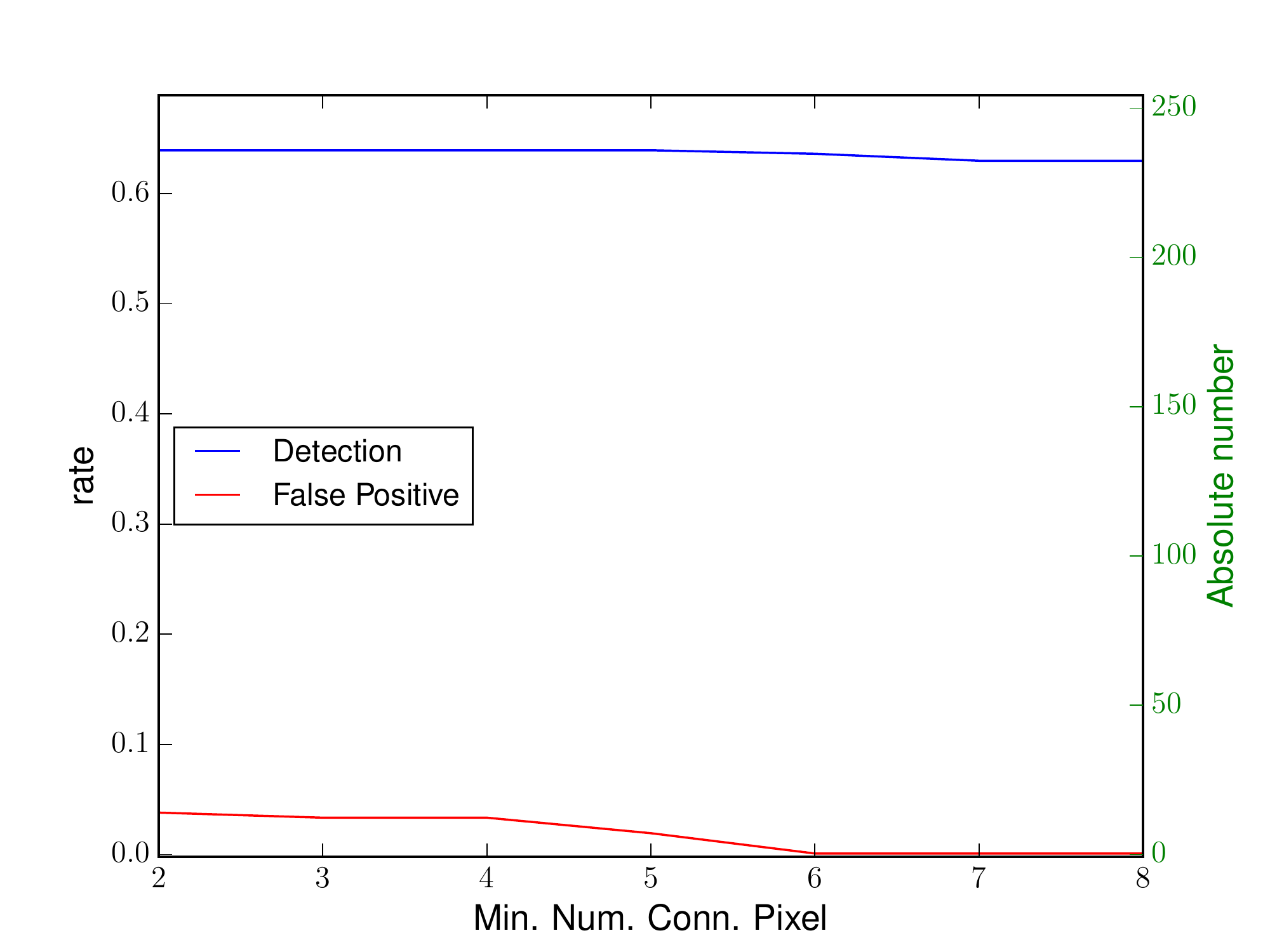}
\caption{\small \label{optimize_ex1}
Detection (blue line) and false positive (red line) rates as a function of the minimum number of connected pixels (setup containing 369 PS with flat $S dN/dS$ and random positions). The plot refers to the 30 –- 100 MeV band.}
\end{figure}

\begin{figure}[h]
\centering
\includegraphics[scale=\onepicabs]{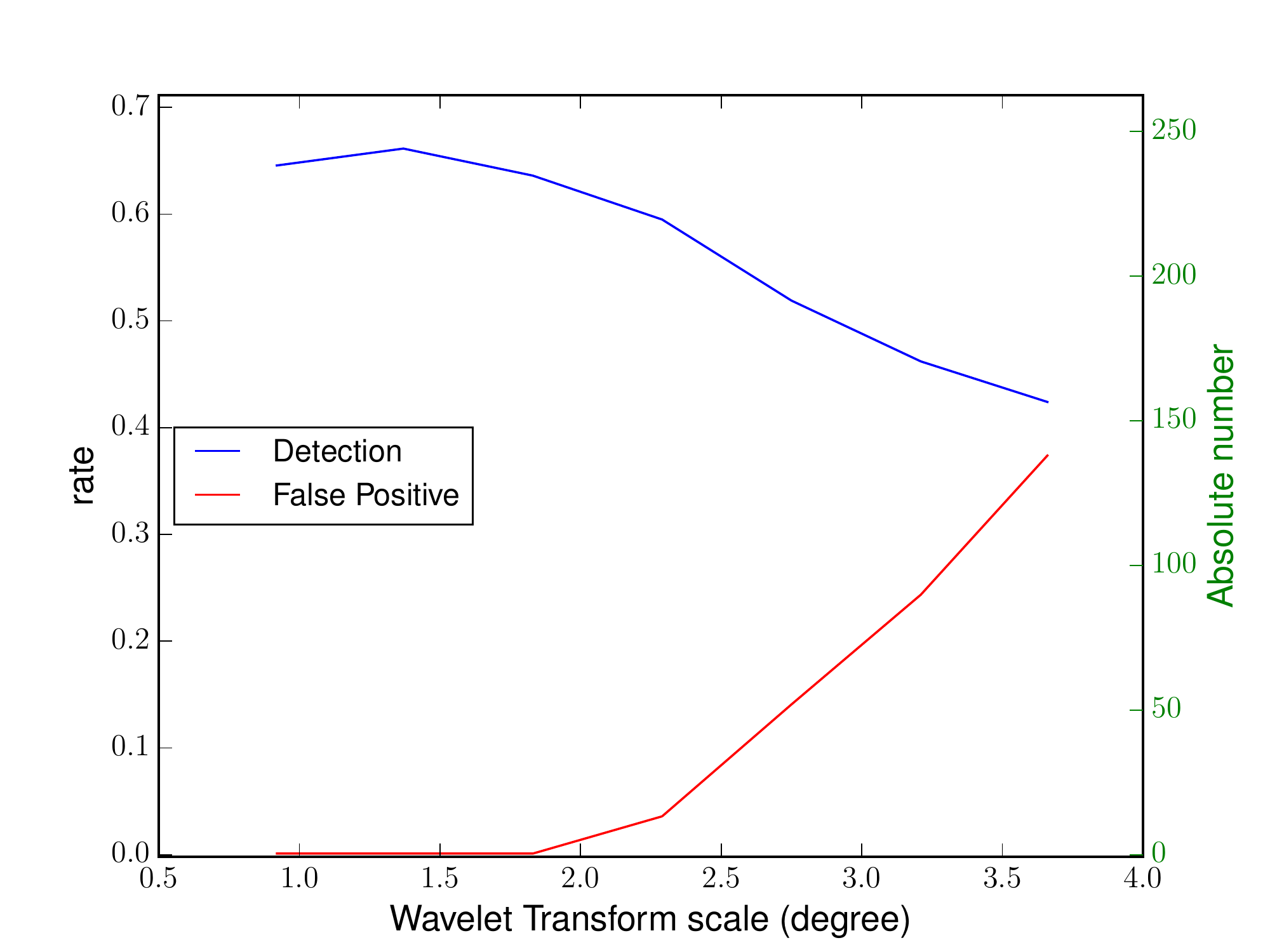}
\caption{\small
\label{optimize_ex2}
Detection (blue line) and false positive (red line) rates as a function of the wavelet transform scale (setup containing 369 PS with flat $S dN/dS$ and random positions). The plot refers to the 30 –- 100 MeV band.}
\end{figure}

Figure \ref{optimize_ex1} shows an example of behavior of detection and false positive rates due to the variation of the minimum number of connected pixels,
while Figure \ref{optimize_ex2} shows the rates as a function of the wavelet transform scale.

\begin{figure}[h]
\centering
\includegraphics[scale=\onepicabs]{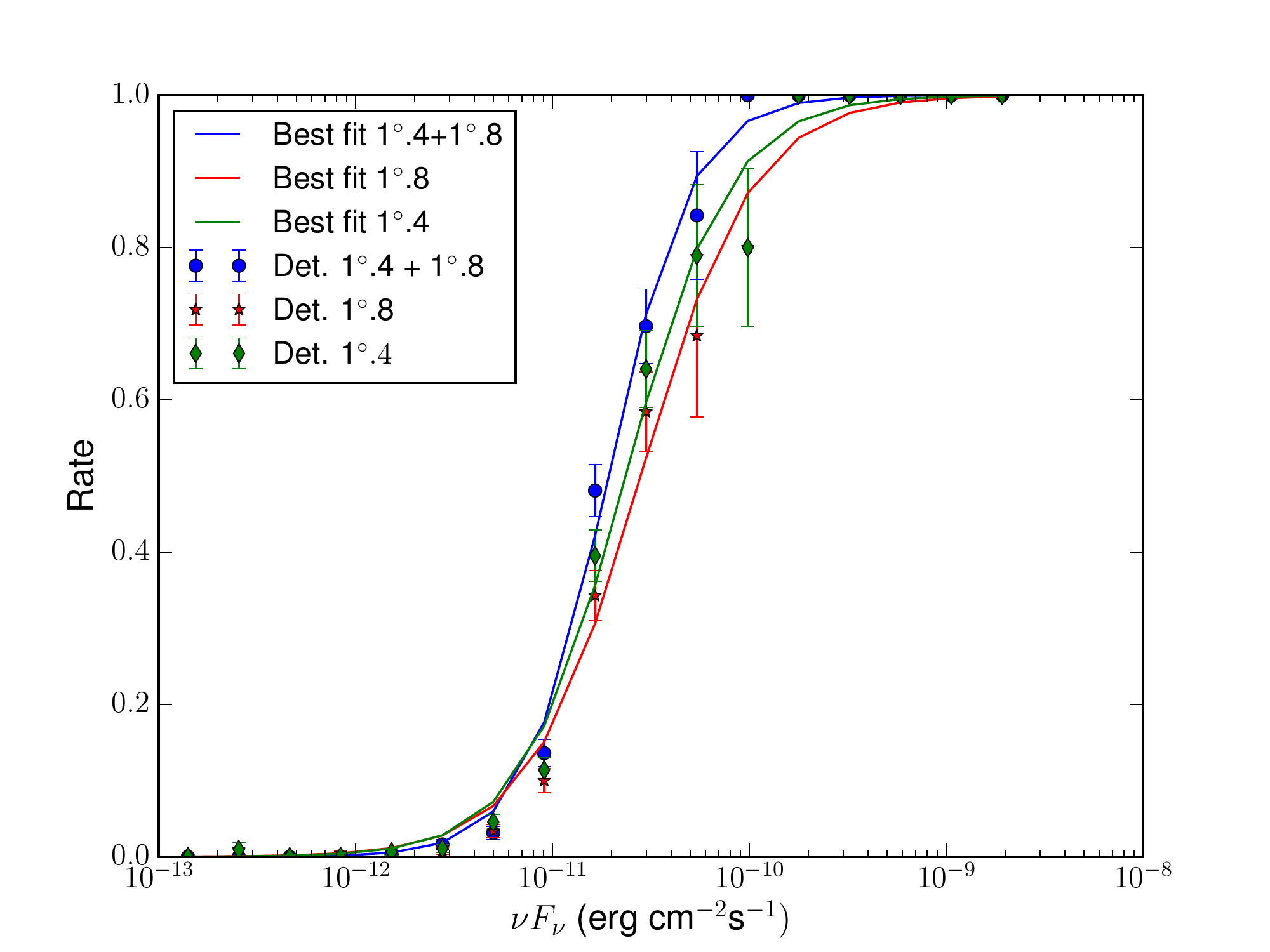}
\caption{\small \label{detection_flux_comb}
Detection rate as a function of the input flux of the simulated PS with random position in the sky (setup containing extrapolated 3FGL sources with randomized position) for high latitude sources, namely $\mid b \mid > 10 \adeg$. 
Each point represents the ratio of detected sources using PGWave with respect to the number of input sources with a flux value inside the flux-bin.
In blue the detection rate of the combined analysis (wavelet scales of 1\fdg4 and 1\fdg8), in red the same for the analysis with a wavelet scale of 1\fdg8 only and in green for that one with a wavelet scale of 1\fdg4 only. }
\end{figure}

We chose to use in the analysis of the data the PGWave parameters that maximize the detection rate keeping the false positive rate smaller than 3\% (similar to the 3FGL false positive rate).
The optimal values of the parameters are reported in Table \ref{pgw_final_par}.
The same parameters are observed to be optimal also for simulated maps with sources with a flux extrapolated from the 3FGL and randomly positioned in the sky.
For the analysis of the data, we chose to use two different wavelet scales: 1\fdg4 (important to resolve source confusion, especially in the Galactic plane) 
and  1\fdg8 (important to detect faint sources at high latitudes).
The expected number of spurious sources is equal to 5 in the energy bin 30 -- 100 MeV (17 in the energy bin 100 -- 300 MeV), so the total false positive rate is lower than 2\% (6\%).
Figure \ref{detection_flux_comb} shows the difference in detection rate between the analysis combining two different wavelet scales and the same for the 1\fdg8 wavelet scale only and 1\fdg4 wavelet scale only.

\begin{table*}[h]
\centering
\begin{tabular}{|c|c|}
\hline
PGWave parameter & Chosen value\\
\hline
\hline
MH wavelet transform scale &  1\fdg8 (1\fdg4) \\
\hline
Min number of connected pixels & 6 (5)\\
\hline
Min distance between sources & 2\fdg7 (2\fdg3)\\
\hline
\end{tabular}
\caption{ \label{pgw_final_par} List of PGWave parameters resulting from the optimization. We use these values for the analysis of the data. In parenthesis the PGWave values of the second choices for the combined analysis.}
\end{table*}

\subsection{Optimized Localization}
\label{Optimized_Localization}

\begin{figure}[h]
\centering
\includegraphics[scale=\onepicabs]{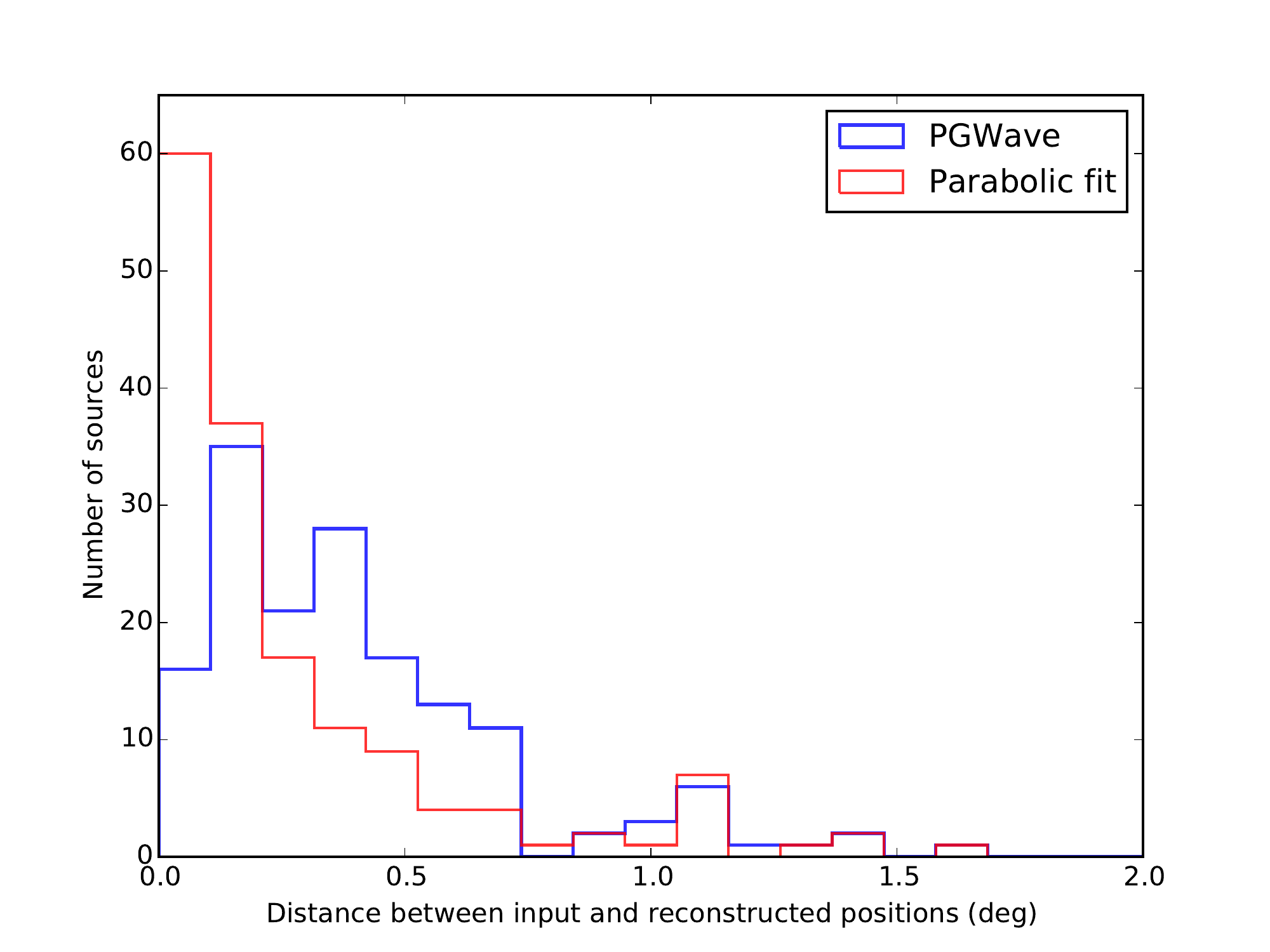}
\caption{\small \label{optmised_position}
Histogram of the distance between the input position and the reconstructed position determined by PGWave, or by the parabolic function fit. We use for the plot the simulated map with ''random'' setup.}
\end{figure}

The PGWave tool determines the position of the sources as the center of the pixel where the WT has a maximum. This reconstruction is limited by the dimension of the pixel (0\fdg458).  We optimize the source localization fitting a two-dimensional parabolic function on the wavelet-transformed map in a 5$\times$5 pixel grid around the maximum.
We correct the PGWave position only for the cases where the fit converges inside the 5$\times$5 pixel grid, for the other cases we keep the position directly determined by PGWave.
This optimization allows improving the localization and reducing the systematic uncertainty.
Figure \ref{optmised_position} shows the distance between input and reconstructed position by PGWave or the reconstructed position where we fit the parabolic function on the wavelet transformed map.
98\% of the sources have a reconstructed position that is localized at less than 1\fdg5 from the input position.
We use this value for the tolerance radius in the association algorithm (see Section \ref{association_algorithm}).

\subsection{PGWave significance }
\label{significance_paragraph}

For the determination of the PS significance, we used MC maps that include diffuse emission and PS with random positions in the sky 
(details on simulated maps can be found in Appendix \ref{appendix_mc_simulations}).
The spectra of the input PS are determined from extrapolation of the spectra of PS randomly selected from the 3FGL catalog.
We apply the analysis to the simulated maps and we compare the resulting sources with the list of input sources in order to estimate the correlation between the PS significance and the input MC flux for each PS.

\begin{figure}[h]
\centering
\hspace{-3mm}
\includegraphics[scale=\onepicabs]{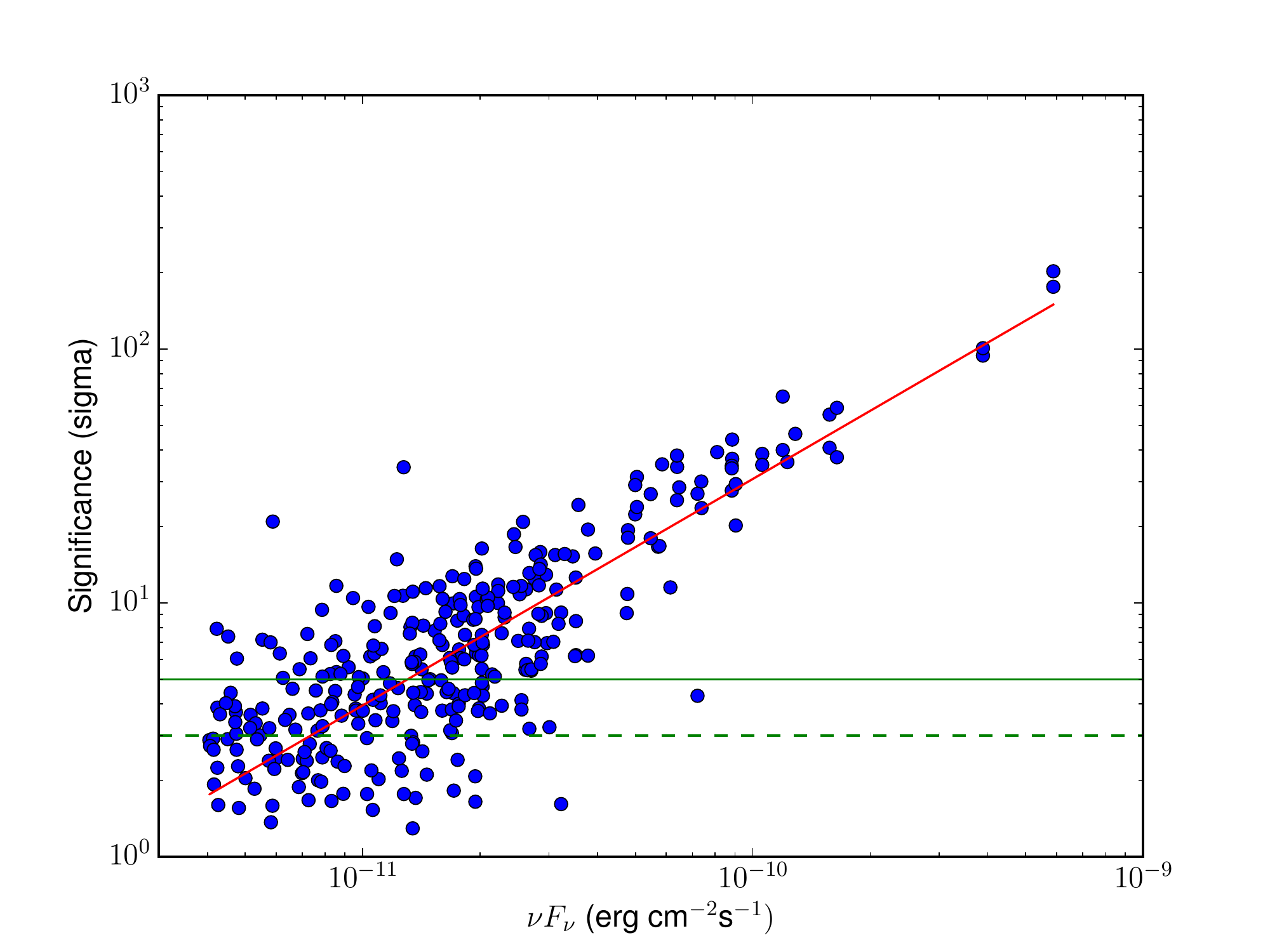}
\caption{\small \label{significance}
PGWave significance vs input MC flux (setup containing extrapolated 3FGL sources with randomized position). The red line represents the best fit with a power-law function ($f = k x^{\alpha}$). In green: with a continues line, is shown the 5 sigma significance used for estimating the statistical sensitivity, instead with a dashed line is shown the 3 sigma significance threshold applied for the point source detection.
}
\end{figure}

Figure \ref{significance} shows, for each seed detected with PGWave, the correlation between the input MC flux and the significance estimated by the wavelets tool. Using a power-law function ($f = k x^{\alpha}$) to fit the data and the obtained fit parameters ($k = 2.6 \times 10^{10}$ and $\alpha=0.9$),
we find that PGWave has a significance greater than 5$\sigma$ for PS with a energy flux $\nu F_{\nu}> 1.3 \times 10^{-11}$ erg cm$^{-2}$ s$^{-1}$.
We use this value as an estimate of the statistical sensitivity of the method in Figure \ref{sensitivity_1fle}.

%\end{appendix}

\end{document}